\documentstyle[aps,epsf,eqsecnum]{revtex}
\tightenlines

 \def\scri{\hbox{${\cal J}$\kern -.645em {\raise
      .57ex\hbox{$\scriptscriptstyle (\ $}}}}

\newcommand{\qed}{\hfill$\Box$}
\newcommand{\proof}{\noindent{\bf Proof:}\ }

\newcommand{\remark}{\noindent{\bf Remark:}\ }

\newtheorem{Theorem}   {Theorem}   [section]
\newtheorem{Corollary} [Theorem]   {Corollary}
\newtheorem{Lemma}     [Theorem]   {Lemma}
\newtheorem{Proposition} [Theorem] {Proposition}
\newtheorem{Conjecture} [Theorem]   {Conjecture}
\begin{document}
\title{Topological Appearance of Event Horizon \\
---What Is the Topology of the Event Horizon\\
 That We Can See?---}

\author{Masaru {\sc Siino}
\footnote{JSPS fellow.} \\
Department of Physics, Kyoto University, Kyoto 606-01}
\maketitle
\centerline{Published in Progress of Theoretical Physics Volume 99, 
    Number 1. pp.1-32 January 1998}
\abstract{The topology of the event horizon (TOEH) is usually believed to be 
a sphere. Nevertheless, some numerical simulations of
  gravitational collapse with a toroidal event horizon or the collision of
  event horizons are reported. Considering the indifferentiability of the 
  event horizon (EH), we see that such non-trivial TOEHs are
  caused by the set of endpoints (the crease set) of the EH.
  The two-dimensional (one-dimensional) crease set is related to 
  the toroidal EH (the coalescence of the EH). Furthermore,
  examining the stability of the structure of the endpoints, it becomes clear
  that the spherical TOEH is unstable under linear perturbation. On the
  other hand, a 
  discussion based on catastrophe theory reveals that the TOEH with
  handles is stable and generic. Also, the relation between the TOEH and 
  the hoop conjecture is discussed. It is shown that the Kastor-Traschen 
  solution is regarded as a good example of the hoop conjecture by 
  the discussion of its TOEH. We further conjecture that a non-trivial 
  TOEH can be smoothed out by rough observation in its mass scale.}

\section{Introduction}
The existence of an event horizon is one of the most
essential concepts in general relativity. As general relativity is 
the theory of spacetime dynamics, it informs us about the causal structure
of a spacetime; there are several kinds of horizons which are the limits
of our communication or influence. In studying the fate of stars, the
event horizon (EH) is the most important among them as the surface of 
a black hole.

To reveal the appearance of a black hole, many authors have been
studying the geometrical properties of the EH. In particular, the area
theorem of an EH\cite{HA} is a noteworthy result, and
its importance in relationship to the properties of black hole entropy is being increasingly
recognized.

Nevertheless, this is not the whole of the EH. It has both
more abstract and more detailed properties. For example, we might be
interested in concrete figures of the EH, e.g., the ellipticity of the 
EH or the multipole moments of its pulsation.  Now, what we wish to 
emphasize is that consideration of the EH from the viewpoint of geometrical
concepts is always based upon more abstract concepts. So, we must
concentrate on the abstract concept of the geometry
of the EH.

The (spatial) topology of an EH is the most fundamental
concept  when one
investigates the various physical properties of the EH, and it is sometimes
believed to be 
trivial. For example, one may assume that the 
topology of each component of the EH is a sphere for the 
uniqueness theorem of a black hole. On the other 
hand, it is natural for the topology of the event horizon (TOEH) to be a
sphere, in an 
astrophysical sense since it is just the surface of a black hole.

So, are the possibilities of a non-spherical TOEH denied?
There is a numerical work reporting the existence of an EH
with non-spherical topology.\cite{ST} After all, attempts to prove 
the TOEH must be a 
sphere in general, are far from perfect.   
In this situation, it might be more interesting to reveal the mechanism 
of   the TOEH than to prove the spherical nature of the TOEH.
In the present article we consider this problem.

In a physically realistic gravitational collapse, it is believed that 
a spacetime is quasi-stationary far in the future. Then, it is natural to 
assume that the TOEH should be a sphere for a single
 asymptotic region, in such a quasi-stationary phase.\cite{HA}\cite{CW}
Supposing that the TOEH is a sphere far in the future, the problem of
the TOEH turns out that of a topology change in a
three-dimensional null surface, i.e., from a two-surface with an arbitrary 
topology to a 
single sphere. From this viewpoint we adapt the well-known theories of the
topology change of a spacetime to this case. Our investigation will show the relation 
between the TOEH and the endpoints (the crease set) of the EH.

Understanding the mechanism of a non-trivial TOEH, the question as to what
topology is probable in general arises. To get a final answer to this
we should determine the whole dynamics of spacetimes. Though 
numerical simulations might be achieve this, we will examine
the stability
or generality of the TOEH to stress something with regard to this question. In a
background of the Oppenheimer-Snyder spacetime, the stability of the
TOEH under linear perturbation is studied. Furthermore, we discuss the
structural stability of the TOEH in the context of catastrophe theory. 
Moreover, we obtain some information about the EH of the
Kastor-Traschen (KT) solution numerically. Since the KT solution
describes the merging of some spherical EHs, we analyze this solution as 
an example of a topologically changing EH. According to our theories of
the TOEH, it is shown that the solution can be regarded as a good 
example of the hoop conjecture of an EH. We also state a conjecture 
concerning the scale dependence of the TOEH assuming a  ``hoop theorem''.

In the next section, we briefly summary  previous works done by other
authors regarding the TOEH. In \S 3 the theorems of topology change are
prepared, and it is applied to the EH\cite{MS1} in \S 4. Section 5 gives 
discussion of linear perturbation in a 
spherically symmetric spacetime.\cite{MS2} In \S 6, the structural 
stability of the TOEH is investigated,\cite{MS2} based on catastrophe 
theory. We study the relation between an example of the exact solution with the change of the
TOEH (the KT solution) and the hoop conjecture, and give a conjecture 
concerned with it\cite{ID} in \S
7. The final section is devoted to summary and discussion.

\section{Background}
The existence of an EH is the most
interesting concept of the causal structure of a spacetime. Many 
authors have studied  
the EH.
Mathematically, the EH is defined as the boundary of the
 causal past of the future
null infinity.\cite{HA} Since the natural asymptotic structure of
a spacetime is supposed to be
  asymptotically flat, where the topology of the future null infinity is 
  $S^2\times R$,
   we naively think that the (spatial) topology of the EH
    will always be $S^2$.

Earlier studies of exact black hole solutions, e.g., the Schwarzschild
solution, Kerr solution, etc., found spherical EHs.
The first work generally dealing with the TOEH is due to Hawking in
1972.\cite{HA} In his work, it is proved in a stationary spacetime that
each component of any (smooth) EH must have a spherical topology. Its extension
to non-stationary spacetimes was first attempted by Gannon.\cite{GA} With the
physically reasonable conditions of the asymptotic flatness and the dominant 
energy condition, it is proved that the topology of a smooth EH must be 
either a sphere or a torus. In 1993, a new lead to investigate the TOEH, 
the ``topological censorship'' theorem, was proven by Friedmann, Schleich and 
Witt.\cite{FS} Assuming asymptotic flatness, global hyperbolicity, and
a suitable energy condition, that theorem stipulates that any two causal
curves extending from the past to the future null infinity are homotopy
equivalent to each other in the sense of a continuous map preserving
their causality. This suggests that the region outside  black holes is
simply connected. From this, Browdy and Galloway concluded that if no
new null generators enter the horizon at later times, the topology of an
EH must eventually be spherical.\cite{BG} This theorem was
strengthened by Jacobson and Venkataramani, limiting the time for which
a torus EH can persist.

For all these works restricting the possibilities of the TOEH, it is  
assumed that the EH is differentiable or there is no endpoint of the EH 
implying the indifferentiability of the EH in a given region. Nevertheless there is no a 
priori reason for analyticity.
 Considering the 
entire structure of an EH, it cannot be 
differentiable in general. For example, even in the case of a spherically symmetric 
spacetime (Oppenheimer-Snyder spacetime) the EH is 
not differentiable where the EH is formed. For stationary black holes, Chru\'{s}ciel and Wald pointed 
out that the toroidal topology of black holes ---as well 
as all other non-spherical topologies--- can be excluded as a simple 
consequence of the ``topological censorship'' theorem, when a suitable 
energy condition is imposed, without assuming the differentiability 
conditions on the EH implicitly assumed in Ref.~\cite{HA}. On the other 
hand, there is no attempt at dealing with indifferentiable dynamical EHs 
like the above examplified formation of the EH. If the 
EH is not smooth, we cannot say that such an EH should generally be a sphere.
    
In fact, the existence of an EH whose topology is not a 
single $S^2$ is 
reported in the numerical 
simulations of gravitational collapse. Shapiro, and Teukolsky et al.\cite{ST}
numerically discovered a toroidal EH in the collapse of toroidal 
matter, and Seidel et al. numerically showed the 
coalescence 
of two spherical EHs.\cite{NCSA} 
This is because, as it will be shown in the present paper, an EH is not 
differentiable at the endpoints of the null geodesic generating the EH.
 In the 
present  paper, we are mainly concerned with such an indifferentiability at the
endpoints. We are not concerned with indifferentiability 
not caused by the endpoints (for example, indifferentiability related 
to the pathological structure of the null infinity\cite{CG}).

\section{The topology change of a spacetime}
Many works have been concerned with the topology change of a spacetime. Some of 
these are useful to discuss the topology of an EH which
is a three-dimensional null surface imbedded into a four-dimensional
spacetime.
 Now we 
briefly present several theorems concerning the topology change of a spacetime.
\subsection{The Poincar\'{e}-Hopf theorem}
Our investigation is based on a well-known theorem regarding the relation 
between the topology of a manifold and a vector field on it.\footnote{It should 
be noted that we never take the affine parametrization of a vector field 
so that the vector field is continuous even at the endpoint of the curve tangent 
to the vector field since we deal with the endpoint as the zero of the vector 
field. If we chose affine parameters, the 
vector field is not unique at the crease set (see next section).}
The following
Poincar\'{e}-Hopf theorem (Milnor 1965) is essential for our investigation.
\begin{Theorem}{\bf Poincar\'{e}-Hopf}\label{TP}
             Let $M$ be a compact $n$-dimensional $(n\ge 2)$ $C^r(r\ge1)$ manifold. 
        $X$ is 
        any $C^{r-1}$ vector field with at most a finite number of zeros, 
        satisfying the following two 
        conditions:  (a) The zeros of $X$ are contained 
        in $Int M$. (b) $X$ has outward directions at $\partial M$. 
        Then the sum of the indices of $X$ at all its zeros is equal to the 
        Euler number $\chi$ 
        of $M$:
        \begin{equation}
                        \chi(M)=ind(X).
                \label{}
        \end{equation}
\end{Theorem}

The index of the vector field $X$ at a zero $p$ is defined as follows. 
Let $X_a(x)$ be the components of $X$ with respect to local coordinates 
$\{x^a\}$ in a neighborhood about $p$. Set $v_a(x)=X_a(x)/\vert X\vert$. If 
we evaluate $v$ on a small sphere centered at $x(p)$, we can regard 
$v_a(S^{n-1})$ as a continuous mapping\footnote{For the theorem in this
 statement, we only need a 
continuous vector field and the index of its zero defined by the 
continuous map $v:S^{n-1}\rightarrow S^{n-1}$. Nevertheless, if one wishes to 
relate the index and the Hesse matrix $H=\nabla_av_b$, a $C^2$ manifold and 
a $C^1$ vector field will be required.} from $S^{n-1}$ into 
$S^{n-1}$. 
The mapping degree of this map is called the index of $X$ at the zero $p$.
For example, if the map is homeomorphic, the mapping degree of the orientation 
preserving 
(reversing) map is $+1$ ($-1$). Figure \ref{fig:zeros} gives some examples of the zeros in two 
and three dimensions.

In the present article, we treat a three-dimensional manifold imbedded into 
a four-dimensional spacetime manifold as an EH. The three-dimensional manifold 
has two two-dimensional boundaries as an initial 
boundary and a final boundary (which is assumed to be a sphere in the next
 section). For such a manifold, we use the following 
modification of the Poincar\'{e}-Hopf theorem. 
Now we consider an odd-dimensional manifold with two boundaries, 
$\Sigma_{1}$ and $\Sigma_{2}$. 

\begin{Theorem}{\bf Sorkin 1986}\label{TS}
                Let $M$ be a compact {\it n}-dimensional ($n > 2$ is 
                an odd number) $C^r(r\ge 1)$ manifold 
                with $\Sigma_1\cup\Sigma_2=\partial M$ and 
                $\Sigma_1\cap\Sigma_2=\phi$. $X$ is 
        any $C^{r-1}$ vector field with at most a finite number of zeros, 
        satisfying the following two 
        conditions: (a) The zeros of $X$ are contained 
        in $Int M$. (b) $X$ has inward directions at $\Sigma_1$ and outward 
        directions at $\Sigma_2$. 
        Then the sum of the indices of $X$ at all its zeros is related to 
        the Euler numbers of $\Sigma_1$ and $\Sigma_2$:
        \begin{equation}
                \chi(\Sigma_{2})-\chi(\Sigma_{1})=2 ind(X).
                \label{eqn:ss}
        \end{equation}
\end{Theorem}

A proof of this theorem is given in Sorkin's work.\cite{SR}

\subsection{Geroch's theorem}
Geroch proved that there is no topology change of a spacetime without a
closed timelike curve.\footnote{Originally he assumed a $C^{\infty}$-differentiable 
spacetime. Nevertheless, his theorem is easily generalized to a $C^r(r\ge 
2)$ 
spacetime.}\cite{GR}

\begin{Theorem}{\bf Geroch 1967}\label{TG}
Let $M$ be a $C^r(r\ge 2)$ $n$-dimensional compact spacetime manifold whose boundary is the 
disjoint union of two compact spacelike $(n-1)$-manifolds, $\Sigma_1$ and 
$\Sigma_2$. Suppose $M$ is isochronous and has no closed timelike curve. 
Then $\Sigma_1$ and $\Sigma_2$ are $C^{r-1}$-diffeomorphic, and $M$ is 
topologically $\Sigma_1\times [0,1]$. 
\end{Theorem}

This theorem is not directly applicable to a null surface $H$, where
 a chronology is determined by
null geodesics generated by a null vector field $K$. In this case, ``isochronous''
means that there is no zero of $K$ in the interior of $H$. On the other hand, the closed
 timelike curve
does not correspond to a closed null curve in a rigorous sense, since on a null surface
an imprisoned null geodesic cannot be distorted, remaining
null, so as to
 become a closed curve as stipulated by Theorem \ref{TG}.\cite{GR}
Then we require a strongly causal condition\cite{WA} on a spacetime
 rather than the condition of no closed causal
curve. The following modified version of Geroch's theorem
arises.
\begin{Theorem}\label{TG2}
Let $H$ be a $C^r (r\ge 2)$ $n$-dimensional compact null surface whose boundary
is the disjoint union of two compact spacelike $(n-1)$-manifolds,
$\Sigma_1$ and $\Sigma_2$. Suppose that there exists a $C^{r-1}$ null vector
 field $K$ which is
nowhere zero in the interior of $H$ and has inward and outward directions at $\Sigma_1$ and
$\Sigma_2$, respectively, and $H$ is 
imbedded into a strongly causal spacetime $(M,g)$. Then $\Sigma_1$ and
$\Sigma_2$ are $C^{r-1}$-diffeomorphic, and $H$ is topologically
$\Sigma_1\times [0,1]$.
\end{Theorem}
\proof
Let $\gamma$ be a curve in $H$, beginning
on $\Sigma_1$, and everywhere tangent to $K$. Suppose first that
$\gamma$ has no future endpoint both in the interior of $H$ and its
boundary $\Sigma_2$. Parametrizing $\gamma$ by a continuous variable $t$ 
with range zero to infinity, the infinite sequence $P_i=\gamma(i),
\ i=1,2,3,\cdots$, on the compact set $H$ has a limit point $P$. Then for any 
positive number $s$, there must be a $t>s$ with $\gamma(t)$ in a
sufficiently small open neighborhood ${\cal U}_P$ (since $P$ is a limit
point of $P_i$), and a $t' > s$ with $\gamma(t')$ not in ${\cal
  U}_P$ (since $\gamma$ has no future endpoint). That is, $\gamma$ must
pass into and then out of the neighborhood ${\cal U}_P$ an infinite
number of times. Since ${\cal U}_P$ can be regarded as the open
neighborhood of $\gamma(t)\in {\cal U}_P$, this possibility is
excluded by the hypothesis that $H$ is imbedded into a strongly causal
spacetime $(M,g)$. Then such a curve $\gamma$ must have a future endpoint
on $\Sigma_2$, because there is no zero of $K$ which is the future endpoint of 
$\gamma$ in the interior of
$H$, from the assumption of the theorem. Hence we can draw the curve $\gamma$ through each point $p$ of
$H$ from $\Sigma_1$ to $\Sigma_2$. By defining the appropriate parameter 
of each $\gamma$, the one
parameter family of surfaces from $\Sigma_1$ to $\Sigma_2$ passing
thorough every point of $H$ is given.\cite{GR} Furthermore the $C^{r-1}$-congruence $K$
provides a one-one correspondence between any two surfaces of this
family. Hence, $\Sigma_1$ and $\Sigma_2$ are $C^{r-1}$-diffeomorphic and
$H\sim\Sigma_1\times [0,1]$.\qed


\section{The topology of event horizon}
Now, we apply the topology change theories given in the previous section to 
EHs.
Let $(M,g)$ be a four-dimensional $C^\infty$ spacetime whose topology is 
$R^4$.
In the remainder of this article, the spacetime  $(M,g)$ is assumed to 
be strongly causal, and also the 
weak cosmic censorship is assumed. Furthermore, 
for simplicity, 
the topology of the EH
(TOEH\footnote{The TOEH is the topology of the spatial section of the EH. 
Of course, it
depends on a timeslicing.}) is assumed to be a smooth $S^2$
far in the future and the EH is not an eternal one (in other words, the EH 
begins somewhere in the spacetime, and it is open to the infinity in the 
future direction with a smooth $S^2$ section).
We expect that those assumptions could be
 valid if we consider only one regular 
($\sim R\times S^2$)
asymptotic region, namely the future null infinity 
$\scri^+$, to define the EH,
and the formation of a black hole. 
The following
investigation, however, is easily extended to the case of different final 
TOEHs far in the future.

In our investigation, the most important concept is the existence of 
the endpoints of
 null geodesics $\lambda$ which lie completely in the EH and generate it. We call these
 the endpoints of the EH.
To generate the EH the null geodesics $\lambda$ are maximally extended to 
the future and past as long as they belong to the EH. Then the endpoint 
is the point where such null geodesics are about to come into the EH (or 
go 
out of the EH), though the null 
geodesic  can continue to the outside or the inside of the EH through 
the    endpoint in the sense of the whole spacetime.
 We consider a null vector field $K$ on the EH which is tangent to the null 
 geodesics $\lambda$.
$K$ is not affinely parametrized, but parametrized so as to be continuous 
even on the endpoint where the caustic of $\lambda$ appears. Then the endpoints of $\lambda$ are the zeros of 
$K$, which can become only past endpoints, since $\lambda$ must reach to 
infinity in the future direction. Of course, using an affine
parametrization, $K$ becomes ill-defined at a subset of the set of the
endpoints. We call such a subset the {\it crease set}. To be
precise, we define the crease set by the set of the endpoints contained
by two or more null generators of the EH. Thus the set of the endpoints consists
of the crease set and endpoints contained by one null generator. As
stated in Ref. \cite{BK}, the crease set contains the interior of the
set of the endpoints and the closure of the crease set contains the set
of the
 endpoints.\footnote{Though Ref. \cite{BK}
  deal with a Cauchy horizon, the same proof is available for an EH.}

Moreover, the fact that the EH defined by 
$\dot{J^-}(\scri^+)$ (the boundary of the causal past of the future null 
infinity) is an achronal boundary (the boundary of a future set) tells 
us that the EH is 
an imbedded $C^{1-}$ submanifold without a boundary (see Ref.~\cite{HE}).
Introducing the normal coordinates $(x^1,x^2,x^3,x^4)$ in a neighborhood 
${\cal U}_\alpha$ about $p$ on the EH, the EH is immersed as 
$x^4=F(x^1,x^2,x^3)$, where $\partial/\partial x^4$ is timelike. Since the 
EH is an achronal boundary, $F$ is a Lipschitz function and one-one map 
$\psi_\alpha:{\cal V}_\alpha\rightarrow R^3,\ \psi_\alpha(p)=x^i(p)$ is a 
homeomorphism, where ${\cal V}_\alpha$ is the intersection of ${\cal 
U}_\alpha$ and the EH\cite{HE}. Then the EH is an imbedded three-dimensional 
$C^{1-}$ submanifold.

First we study the relation between 
the crease set and the differentiability of the EH. We see that
the EH is not differentiable at the crease set.    

\begin{Lemma}\label{L1}
Suppose that $H$ is a three-dimensional null surface imbedded into the spacetime $(M,g)$ by 
a function $F$ as
\begin{equation}
H:x^4=F(x^i,i=1,2,3),
\end{equation}
in a coordinate neighborhood 
$({\cal U}_{\alpha},\phi_\alpha),\ \phi_\alpha:{\cal U}_{\alpha}\rightarrow 
R^4$, where 
$\partial/\partial x^4$ is timelike.
When $H$ is generated by the set of null geodesics  whose tangent vector 
field is $K$, we define the crease set by the set of the endpoints
of the null geodesics contained by two or more null generators of $H$. 
Then, $H$ and the imbedding function $F$ are indifferentiable at the crease set.
\end{Lemma}
\proof
If $H$ is a $C^r(r\ge 1)$ null surface around $p$, we can define the tangent space 
$T_p$ of $H$, which is spanned by one null vector and two independent 
spacelike vectors. On the contrary, the point in the crease set is
contained by two or more null generators of $H$. Therefore, there exists 
two or more null vectors tangent to $H$ at $p$, and there is no unique
choice of a
null vector defining $T_p$. This implies that $H$ and the imbedding
function $F$ is not differentiable at the crease set.\qed

In the present article, we deal only with this indifferentiability.
 Then, 
we assume that the EH is $C^r(r\ge 2)$-differentiable (the inequality
$r\ge 2$ is necessary for Theorem \ref{TG2}), except on the 
crease set of the EH and we assume that the set of the 
endpoints is compact. Thus we suppose that the EH is indifferentiable 
only on a compact subset.
Incidentally, in the case where the future null infinity possesses pathological
structure, the EH could be nowhere differentiable.\cite{CG} 
Nevertheless we have no concrete example of a physically reasonable spacetime 
with such a non-compact indifferentiability. Similarly there might be
the case where an indifferentiable point is not the endpoint of the EH. 
In spite of this possibility, the reason we
 consider only the indifferentiability
caused by the endpoints is that every EH possesses at least one
 endpoint, 
except for eternal EHs. Most of the indifferentiability which we can 
imagine would be concerned with the endpoint. 

Next, we prepare a basic proposition.
Suppose there is no past endpoint of a null geodesic generator of an EH between 
$\Sigma_1$ and $\Sigma_2$. 
 Then, Geroch's theorem stresses  the topology of the smooth
EH does not change. 

\begin{Proposition}  \label{P1}
Let $H$ be a compact subset of the EH of $(M,g)$ whose boundaries are an initial spatial section
$\Sigma_1$ and a final spatial section $\Sigma_2$,
$\Sigma_1\cap\Sigma_2=
\emptyset$. $\Sigma_2$ is assumed to be
a smooth sphere far in the future. Suppose that $H$ is $C^r(r\ge 
2)$-differentiable. Then
the topology of $\Sigma_1$ is $S^2$.
 \end{Proposition}
\proof
As proved in Ref. \cite{BK}, if there is any endpoint of the null geodesic generator of the EH 
in the interior of $H$,
 $H$ cannot be $C^1$-differentiable there. Using Theorem
\ref{TG2}, it is concluded that $\Sigma_1$ is topologically $S^2$, since
$H$ is imbedded into a strongly causal spacetime $(M,g)$.
\qed

Now we discuss the possibilities of non-spherical topologies.
From Sorkin's Theorem there should be at least one zero of 
 $K$ in the interior of $H$ provided that
the Euler number of $\Sigma_1
$ is different from that of $\Sigma_2\sim S^2$. Such a zero  can only be the past
endpoint of the EH, since the null 
geodesic generator of the EH
cannot have a future endpoint. With regard to this past
 endpoint and the crease set of the EH we state 
the following two propositions.

\begin{figure}
\centerline{\epsfxsize=18cm\epsfbox{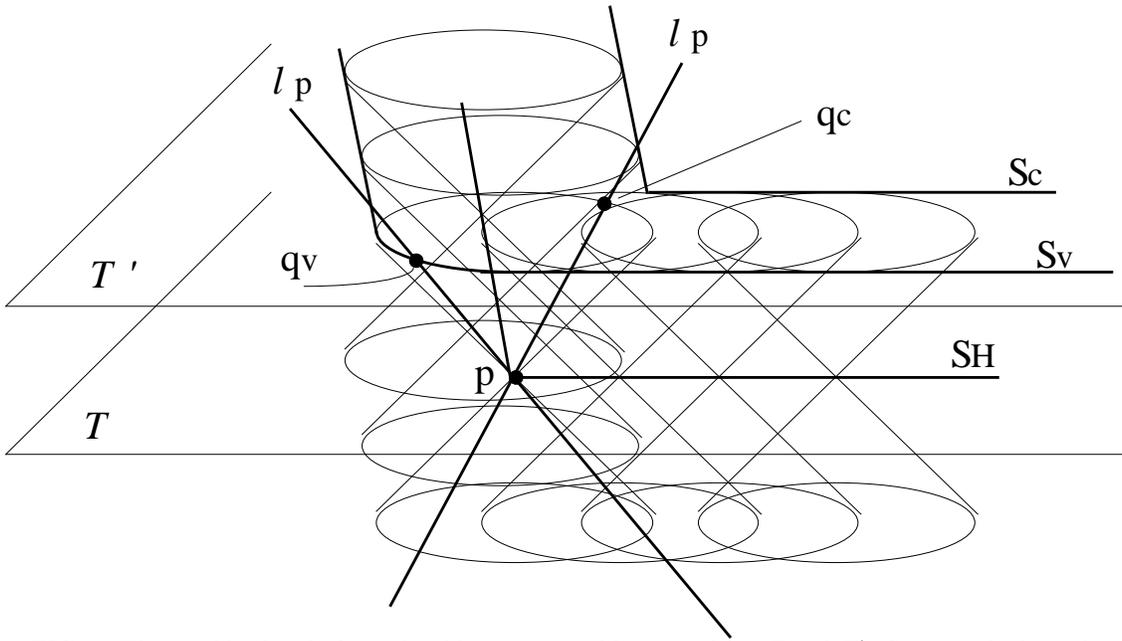}}
\caption{The neighborhood of $p$ is sliced by two spatial hypersurfaces
  $\cal T$ and $\cal T'$.
$S_H$ is on the lower hypersurface $\cal T$. $\ell_p$ passes through $p$. 
In the convex (concave) case, the EH is given by the enveloping surface 
$S_v$ ($S_c$). $q_v$ ($q_c$) is a point on $\ell_p$ at the future of $p$. 
$S_v$ is $C^1$-differentiable at $q_v$. $q_c$ is inside $S_c$.}
\label{fig:cones}
\end{figure}

\begin{Proposition}\label{P2} 
           The crease set (consisting of the past endpoints) 
 of the EH 
is an acausal  set.
\end{Proposition}
\proof
The crease set is obviously an achronal set, as the EH is a 
null surface (an achronal boundary). 
Suppose that the crease set includes a null segment $\ell_p$ through an event $p$. By
Lemma \ref{L1}, the null segment 
$\ell_p$ consists of the indifferentiable points of the EH. The EH, however, is differentiable in the null direction tangent to 
$\ell_p$ at $p$, since $\ell_p$ is smoothly imbedded into the smooth 
spacetime $(M,g)$. Then the
 section $S_H$ of the EH on a spatial hypersurface through 
$p$ is indifferentiable at $p$, as shown in Fig.~\ref{fig:cones}. 
Considering a sufficiently small neighborhood ${\cal U}_p$ about $p$, 
the local causal
structure of ${\cal U}_p$ is 
similar to 
that of Minkowski spacetime, since $(M,g)$ is smooth there. Therefore, when $S_H$ is 
convex at $p$, the EH will be $C^1$-differentiable at $q_v$, which is on
a nearby future of the 
null segment $\ell_p$ (see Fig.~\ref{fig:cones}), because the EH is the outer side of the 
enveloping surface 
of the 
light cones standing along $S_H$ in the neighborhood ${\cal U}_p$ about $p$. 
Nevertheless, from Lemma \ref{L1}, also $q_v \in \ell_p$ cannot be 
smooth in this section.  Also, if $S_H$ is concave, $q_c$ which is on a nearby 
future of $\ell_p$, will invade the inside of the EH and fails to be on the EH (see 
Fig.~\ref{fig:cones}). Thus the crease set 
cannot contain either convex and concave null segment. 
Moreover if two disconnected segments could be connected by a null geodesic, 
a future endpoint of the null geodesic generator would exist. Hence the
crease set is an acausal  set.
\qed

\begin{Proposition} \label{P3}
           The crease set (consisting of past endpoints) of the EH of $(M,g)$ is
           arc-wise connected. Moreover, the collared crease set is 
           topologically $D^3$.
\end{Proposition}
\proof
Consider all the null geodesics $\lambda_{p_e}(\tau)$ emanating from the 
crease set $\{p_e\}$ tangent to the null 
vector field $K$. Since the crease set is the set of zeros of $K$, $p_e$ 
corresponds to $\lambda_{p_e}(-\infty)$. From Proposition \ref{P2}, 
the spacelike section 
$\cal S$ of the EH very close to the crease set $\{p_e\}$ is determined by a 
map $\phi^K$, with a large negative parameter $\Delta \tau$ of the null 
geodesic $\lambda_{p_e}$:
 \begin{eqnarray}
        \phi^K :\{q\in{\cal S}\}&\longrightarrow& \{p_e\}\\
        s.t.\  \lambda_{p_e}(-\infty)&=&p_e,\ \lambda_{p_e}(\Delta\tau)=q.
        \label{}
 \end{eqnarray}
Here, with a sufficiently large negative $\Delta\tau$ ($\rightarrow -\infty$), $K$ has inward directions
 to $H$ at $\cal S$, 
where $H$
is the subset of the EH bounded by $\cal S$ and the final spatial section $\Sigma_2$ 
which is far in the future and is a smooth sphere from the assumption.
By this construction, the entire crease set is wrapped by $\cal S$, and $\cal 
S$ is compact because of the assumption that the set of endpoints is compact. 
$H$ and the crease set are on the opposite side of $\cal S$.
 Therefore there is no endpoint in the interior of $H$. Since $H$ is 
 $C^r(r\ge 2)$-differentiable except on the crease set and compact from the 
 assumption, Proposition \ref{P1} implies that $\cal S$ is homeomorphic to $ 
 \Sigma_2\sim S^2$ and $H$ is topologically $S^2\times [0,1]$. 
 If there were two or more connected components of the crease set, one would 
need the same number of spheres to wrap it with $\cal S$ being sufficiently close
to the 
crease set. However, since $\cal S$ is homeomorphic to a single $S^2$, the crease set 
should be 
arc-wise connected. In other words, the collared crease set is topologically
 $D^3$, because the EH and the crease set are imbedded into $(M,g)$.\qed

 The set of past endpoints is also arc-wise connected, since the crease
 set is contained by it, and the closure of the crease set contains it.\cite{BK}

Now we give theorems and corollaries regarding the topology of the 
spatial section of the EH on a timeslicing.
First we consider the case where the EH has simple structure.

\begin{Theorem}\label{T1}
Let $S_H$ be the section of an EH 
determined by a spacelike hypersurface.
If the EH is $C^r(r\ge 1)$-differentiable at $S_H$, it is
topologically $\emptyset$ or $S^2$.
\end{Theorem}
\proof
From Lemma \ref{L1}, there is no intersection between $S_H$ and the
crease set.  Since the EH
is assumed not to be eternal, there exists at least one endpoint of the EH in the past of $S_H$
as long as $S_H \neq \emptyset$. Therefore Proposition \ref{P3} implies
there is no endpoint of the EH in the future of $S_H$. By the  
assumption that the EH is $C^r(r\ge 2)$-differentiable except 
on the crease set and Proposition
\ref{P1}, it is concluded that $S_H$ is topologically $S^2$.
\qed

On the other hand, we obtain the following theorem about the change of the  
 TOEH with the aid of Sorkin's
 theorem. Here, we introduce the dimension of the crease
 set. Considering an open subset of the crease set, if the subset is 
 a $n$-dimensional topological submanifold,
we state that the crease set is $n$-dimensional, or the dimension of the 
 crease set is $n$, in the open subset.
Since an
 EH is an imbedded $C^{1-}$ submanifold of a spacetime, the crease set
 where the EH is indifferentiable has three-dimensional measure
 zero.\cite{BK} The crease set is zero-, one- or two-dimensional.
\begin{Theorem}\label{T2}
Consider a smooth timeslicing ${\cal T}={\cal T}(T)$ defined by a smooth 
function $T(p)$:
\begin{equation}
        {\cal T}(T)=\{p\in M\vert T(p)=T=const.,\ \  T\in 
        \left[T_1,T_2\right]\}, \ \ g(\partial_T,\partial_T)<0.
        \label{}
\end{equation}
Let $H$ be the subset of the EH cut by ${\cal T}(T_1)$ and ${\cal T}(T_2)$
whose boundaries are the initial spatial section $\Sigma_1\subset 
{\cal T}(T_1)$ and the final spatial section $\Sigma_2\subset 
{\cal T}(T_2)$, and let $K$ be the null vector field generating the EH.
 Suppose that $\Sigma_2$ is a 
sphere.  If, 
in the timeslicing $\cal T$, the TOEH changes ($\Sigma_1$ is not
homeomorphic to $\Sigma_2$) then there is a crease set (the zeros of $K$) in 
$H$, and when the timeslice touches
\begin{itemize}
        \item the one-dimensional segment of the crease set, it  causes the coalescence of two
        spherical EHs.
        \item  the two-dimensional segment of the crease set, it  causes the change of the TOEH
        from a torus
        to a sphere.
\end{itemize}        
\end{Theorem}
\proof
First of all, we regularize $H$ and $K$ so that Theorem \ref{TS} can be applied 
to this case.
Introducing normal coordinates $(x^1,x^2,x^3,x^4)$ in a neighborhood 
${\cal U}_\alpha$ about $q\in H$,  since 
the EH is an achronal boundary,
$H$ is imbedded by a Lipschitz function $x^4=F(x^i,i=1,2,3)$, where 
$\partial/\partial x^4$ is timelike 
(see Ref. \cite{HE}). Here we set $x^4(p)=T(p)-T(q)$ in ${\cal U}_\alpha$ 
about $q$. Since $M$ is a metric space, there is the partition of unit 
$f_\alpha$ for the atlas 
$\{{\cal U}_\alpha,\phi_\alpha\},\  
\phi_\alpha:{\cal U}_\alpha\rightarrow R^4$.\cite{SS} Then a smoothed function of 
the Lipschitz function $T(p\in H)$ (which is restricted on the indifferentiable 
submanifold $H$ for the smooth function $T(p)$ to become indifferentiable) with a smoothing scale 
$\epsilon$ is given by 
\begin{eqnarray*}
        \widetilde{T}(p\in H)&=&\Sigma_\alpha\int_{\cal{U}_\alpha}f_\alpha T(r=\phi_\alpha^{-1}(x^1,x^2,x^3,x^4)) 
        W(p,r)\delta(x^4-F(x^1,x^2,x^3)) dx^1dx^2dx^3dx^4\\
        &=&\Sigma_\alpha\int_{\cal{U}_\alpha}f_\alpha (F(x^1,x^2,x^3)+T(q)) 
        W(p,r)\delta(x^4-F(x^1,x^2,x^3)) dx^1dx^2dx^3dx^4\\
        W(p,r)&=&0,\ \ \ p\notin {\cal U}_\alpha, \\
        W(p,r)&=&w(     \vert p-r \vert), \ \ \ p\in {\cal U}_\alpha,\\
        \vert p-r \vert&=& 
        \sqrt{(x^1_p-x^1_r)^2+(x^2_p-x^2_r)^2+(x^3_p-x^3_r)^2+(x^4_p-x^4_r)^2} \ \ \ in\ \ {\cal U}_\alpha
         \ \ about\ \ q,\\
        w(x)&\le&\infty,\ \ \ w(x\gg\epsilon)\ll 1,\ \ \ \int w(x)=1,
\end{eqnarray*}
where $w$ is an appropriate window function with a smoothing scale $\epsilon$.
The support of $W$ is a sphere with radii $\sim \epsilon$ and 
$w(\vert x\vert,\epsilon\rightarrow 0)= \delta^4({\bf x})$. Of course, $\epsilon =0$ gives 
the original function $T(p\in H)=\widetilde{T}(p\in H)$. Taking a sufficiently small 
non-vanishing $\epsilon$, a new imbedded submanifold 
$\widetilde{H}$, with $\widetilde{x^4}(p\in \widetilde{H})=\widetilde{T}(p)-
\widetilde{T}(q)=:\widetilde{F}(x^1,x^2,x^3)$ in ${\cal U}_\alpha$ about 
$q$, can 
become homeomorphic to $H$ and $C^r(r\ge 1)$-differentiable. From this 
smoothing procedure, we define a smoothing map $\pi$ (homeomorphism) by;
\begin{eqnarray}
        \pi : H&\longrightarrow& \widetilde{H}\\
\phi_\alpha^{-1}(x^1,x^2,x^3,x^4)&\longrightarrow&\phi_\alpha^{-1}(x^1,x^2,x^3,\widetilde{x^4}).
        \label{}
\end{eqnarray}
Of course, this map depends on the atlas $\{{\cal 
U}_\alpha,\phi_\alpha\}$ introduced.
This smoothing map induces the correspondences
\begin{eqnarray}
\lambda &\longrightarrow& \widetilde{\lambda},\ \ \ 
\Sigma_{1,2}\longrightarrow \widetilde{\Sigma_{1,2}},\\ 
{\cal T} &\longrightarrow& \widetilde{\cal T},\ \ \ 
\pi^*:K\longrightarrow \widetilde{K},
\end{eqnarray}
where $\widetilde{K}$ is the tangent vector field of curves 
$\widetilde{\lambda}$ generating $\widetilde{H}$. 
Now $\widetilde{K}$ is not always null. Hereafter we call also
the image of the crease set by the smoothing map $\pi$ a crease set 
for $\widetilde{\lambda}$,
though the generators $\widetilde{\lambda}$ are not null.

\begin{figure}
\centerline{\epsfxsize=14cm \epsfbox{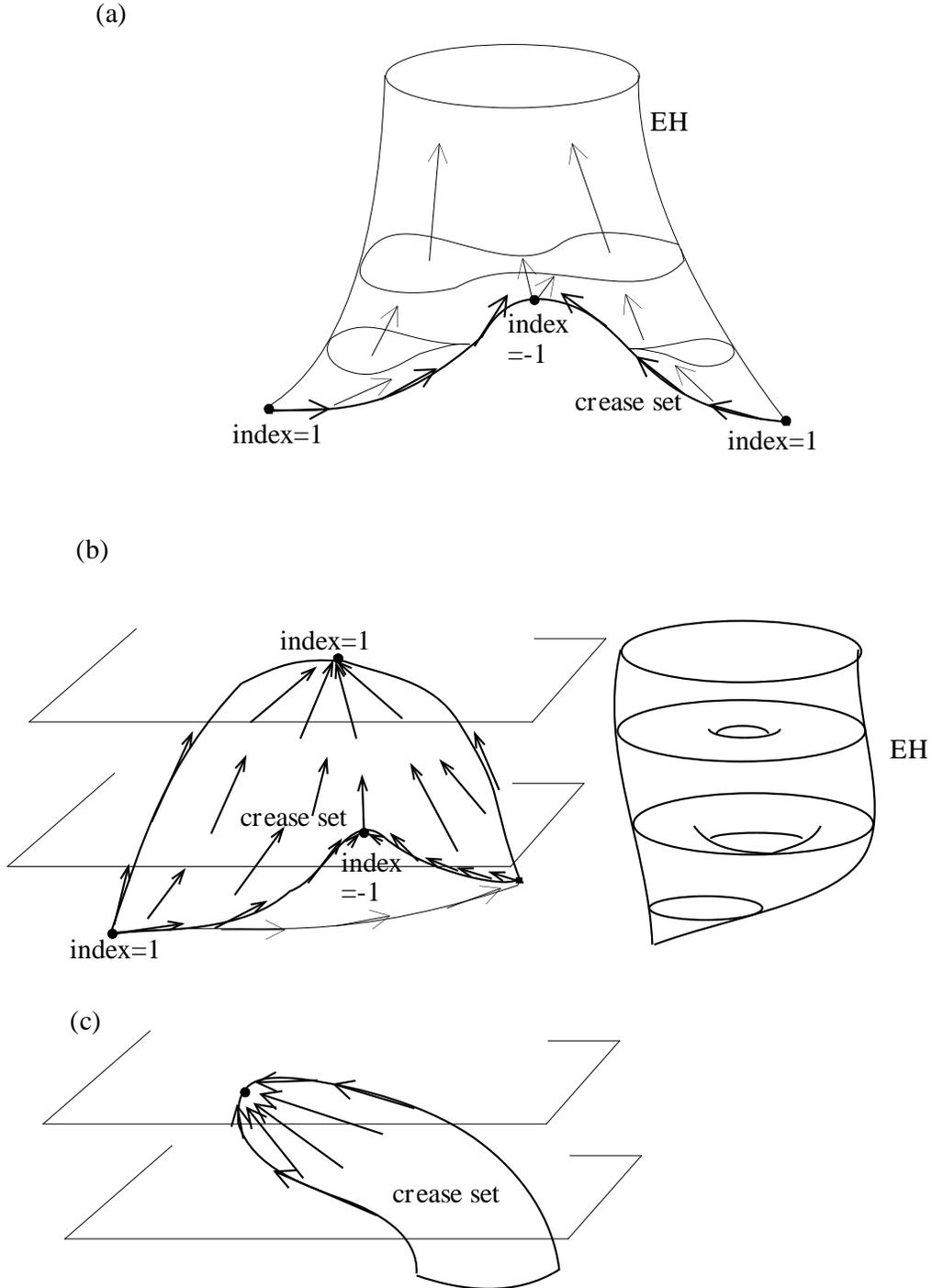}}
\caption{(a) and (b) are the one-dimensional and two-dimensional crease set,
 respectively.  In (b), we draw the entire of the EH separately. (c) is the case
  in which the edge of the crease set is hit from the future.
 By these vector fields $\overline{K}$, 
  the crease sets are generated. The zeros of $\overline{K}$ and their indices are indicated.}
\label{fig:ends}
\end{figure}

Furthermore, using the transformed
timeslicing $\widetilde{\cal T}$, we should modify $\widetilde{K}$ so that 
the crease set for 
$\widetilde{\lambda}$ becomes zero-dimensional, that is, the set
of isolated zeros, (where this set will no longer 
 always be arc-wise connected). To make the zeros isolated, a 
 modified vector field $\overline{K}$ should 
be given on the crease set for $\widetilde{\lambda}$ so as to generate this 
set. 
On the crease set for $\widetilde{\lambda}$, $\overline{K}$ should be determined by 
the timeslice 
$\widetilde{\cal T}$ so that $\overline{K}$ is tangent to the crease set for
$\widetilde{\lambda}$ and 
directed to the future in the sense of the timeslicing $\widetilde{\cal T}$. 
In particular, at the boundary of the crease set for $\widetilde{\lambda}$, 
we should be 
careful that $\overline{K}$ is tangent also to the non-zero-dimensional 
boundary of the crease set for $\widetilde{\lambda}$.
 Here it is noted that the case in which the boundary is 
tangent to the timeslicing $\widetilde{\cal T}$ is possible and we cannot 
determine the direction of $\overline{K}$ there. 
Since such a situation is unstable under the small deformation of the 
timeslicing, however, we omit this possibility, as mentioned in the 
remark appearing after this proof. Hence $\overline{K}$ is determined on 
the crease set for $\widetilde{\lambda}$ (see, for example, Fig.~\ref{fig:ends}) 
and it is the set of some 
isolated zeros. 
At this step, $\overline{K}$ on the crease set for $\widetilde{\lambda}$ and $\widetilde{K}$ (except on 
the set of the zeros for $\widetilde{K}$) is discontinuous. 
Then, we modify $\widetilde{K}$ around the crease set for $\widetilde{\lambda}$
along $\overline{K}$, and make the modified $\widetilde{K}$ into 
$\overline{K}$, except on the crease set for $\widetilde{\lambda}$, 
without changing the characters of the 
zeros, so that $\overline{K}$ becomes a
continuous vector field on $\widetilde{H}$. One may be afraid that an extra 
zero of $\overline{K}$ appears as a result of this continuation. Nevertheless it is 
guaranteed by the existence of the foliation by the timeslice $\cal T$ or 
$\widetilde{\cal T}$ that there exists the desirable modification of 
$\widetilde{K}$ around the crease set for $\widetilde{\lambda}$, since both $\widetilde{K}$ and 
$\overline{K}$ are future directed in the sense of the timeslicing 
$\widetilde{\cal T}$. Thus we get 
$\overline{K}$ and its integral curves 
$\overline{\lambda}$ on the whole of $\widetilde{H}$. 
From this construction of $\overline{K}$, there are 
some isolated zeros of $\overline{K}$ only on the crease set for $\widetilde{\lambda}$,
and
$\overline{K}$ is everywhere 
future directed in the sense of the timeslicing $\widetilde{\cal T}$ 
(though they will be spacelike somewhere). Of course, $\overline{\lambda}$ 
will have both future and past endpoints.

Now we apply Theorem \ref{TS} to $\widetilde{H}$ with the 
modified vector field $\overline{K}$, whose boundaries are 
$\widetilde{\Sigma_1}$ and $\widetilde{\Sigma_2}\sim S^2$. 
Since $\widetilde{\Sigma_1}$ and 
$\widetilde{\Sigma_2}$ are on ${\cal T}(T_1)$ and ${\cal T}(T_2)$, 
respectively, $\overline{K}$ 
has inward directions at $\widetilde{\Sigma_1}$ and outward directions at 
$\widetilde{\Sigma_2}$. 

\begin{figure}
        \centerline{\epsfxsize=14cm \epsfbox{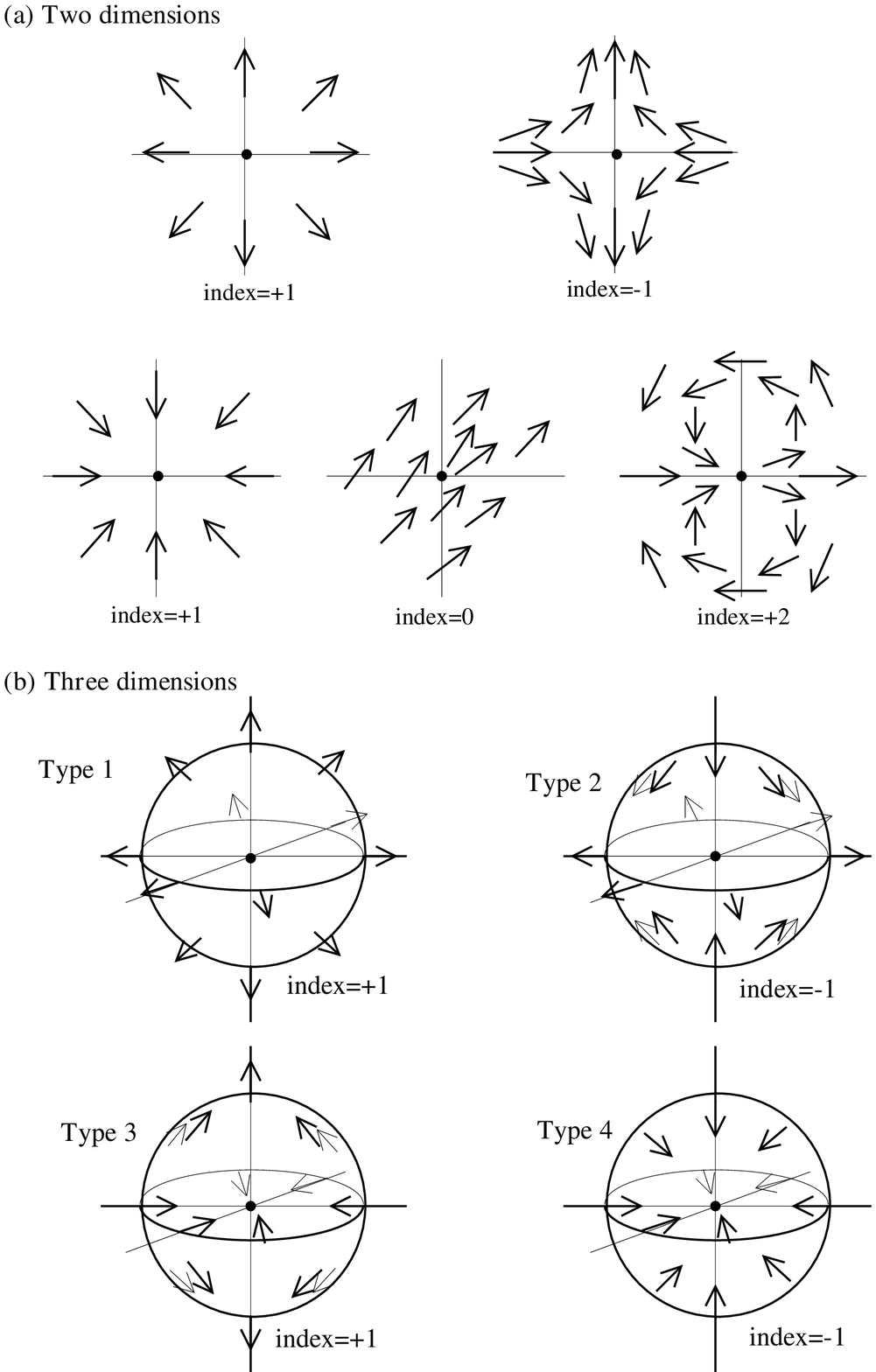}}
        \caption{(a) Two-dimensional zeros and a
 vector field around them.
        Five types of zeros are shown in this figure.
        (b) Three-dimensional zeros and a vector field around them. 
        Only the zeros with $|\rm index |=1$ are shown.
 Other cases can easily
        be understood by analogy to (a).}
        \protect\label{fig:zeros}
\end{figure}

 From the construction above, we see that the type of the zero of $\overline{K}$ depends
on the dimension of the crease set. In particular, for the zero most 
in the future, the one-dimensional crease set
provides the zero of the second type in Fig.~\ref{fig:zeros}(b) corresponding 
to index $=-1$ and the two-dimensional 
crease set gives that of the third type in Fig.~\ref{fig:zeros}(b) with index $=+1$ (see 
Fig.~\ref{fig:ends}). Following Theorem 
\ref{TS}, the Euler number
changes at the zero by an amount $2\times$index. Therefore if there is a
one- (two)-dimensional crease set, the timeslicing $\cal T$ gives the topology change 
of the EH from two spheres (a torus) to a sphere. When $H$ 
contains the whole of the crease set, it will, according to Theorem \ref{TS}, present all changes 
of the TOEH from the formation of the EH to a sphere far in the future, as shown in 
Fig.~\ref{fig:ends}. To complete the discussion, 
we also consider uninteresting cases provided 
by a certain timeslicing. When the edge of the crease set is hit by the  
timeslicing from the future, according to the 
construction above, it gives a zero with its index being zero 
(Fig.~\ref{fig:ends}(c)), and there is no topological change of the EH. \qed

This result is partially suggested in Shapiro, Teukolsky and Winicour.\cite{ST}

\begin{figure}
\centerline{\epsfxsize=18cm \epsfbox{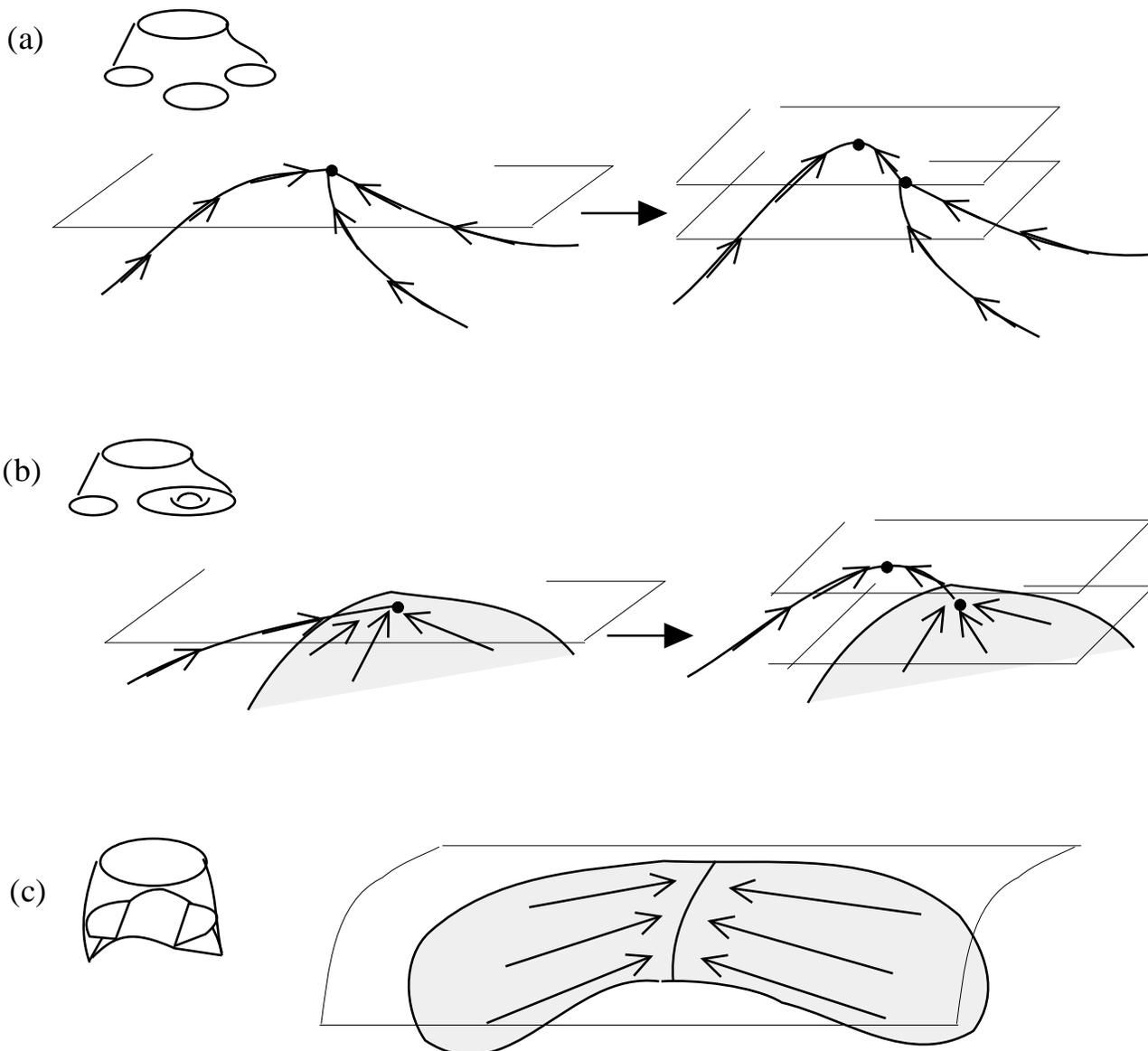}}
\caption{(a) and (b) are the examples of the branching crease set 
in an accidental timeslicing. They are understood by a small 
deformation of the timeslicing. 
On the other hand, (c) is the case in which the timeslicing is 
partially tangent to the crease set.
The two-dimensional crease set behaves as a one-dimensional crease set.}
\label{fig:branches}
\end{figure}

\remark
One may face special situations.
 The possibility of branching 
endpoints should be noted. If the crease set
possesses a branching point, a special timeslicing can make the branching 
point into an isolated zero, though such a timeslicing loses this aspect under 
the small deformation of the timeslicing. The index
of this branching endpoint is hard to be determined in a direct consideration.
 The situation, however, is regarded as the degeneration of the two 
 distinguished zeros of $\overline{K}$ in  
 $\widetilde{H}$. Some examples are displayed in Fig.~\ref{fig:branches}.
 Imagine a slightly slanted timeslicing, and it will decompose the branching point
 into two distinguished zeros(of course, there are the possibilities of the
 degeneration of three or more zeros). The first case is the branch of
  the one-dimensional crease set\footnote{We can also
   treat the branching   points of the two-dimensional crease set in the same
   manner.} (Fig.~\ref{fig:branches}(a)), where
  the branching point is the degeneration of two zeros of $\overline{K}$ 
  with their index
   being $-1$, since they are the results of 
   the one-dimensional 
   crease set. Then the index of the branching point is $-2$ and, for example, three
   spheres coalesce there. The next case is a one-dimensional branch from
    the two-dimensional crease set (Fig.~\ref{fig:branches}(b)).
This branching point is the degeneration of the zeros of $\overline{K}$ from
 the one-dimensional
crease set (index $=-1$) and the two-dimensional crease set (index $=+1$). 
This decomposition reveals that, though the index of this point vanishes, the TOEH changes at
 this point, for example, from a sphere and a torus to a sphere. 
 Of course, the Euler number does not change in this process. 
 Furthermore, these topology changing processes are
 stable under the small deformation of the timeslicing. 
 Finally, there is the case in which a timeslicing is partially tangent to the crease set 
 or its boundary.
 For instance, an accidental timeslicing can hit, not a single point in 
 the crease set, but a curve in the crease set from the future, 
 as shown in Fig.~\ref{fig:branches}(c). For such a timeslicing, the contribution of 
 the two-dimensional crease set to the index 
 is not $-1$ but $1$. This situation, however, is
  unstable under
  the small deformation of the timeslicing, and we omit such a case in the 
  following.

\begin{figure}
        \centerline{\epsfxsize=16cm \epsfbox{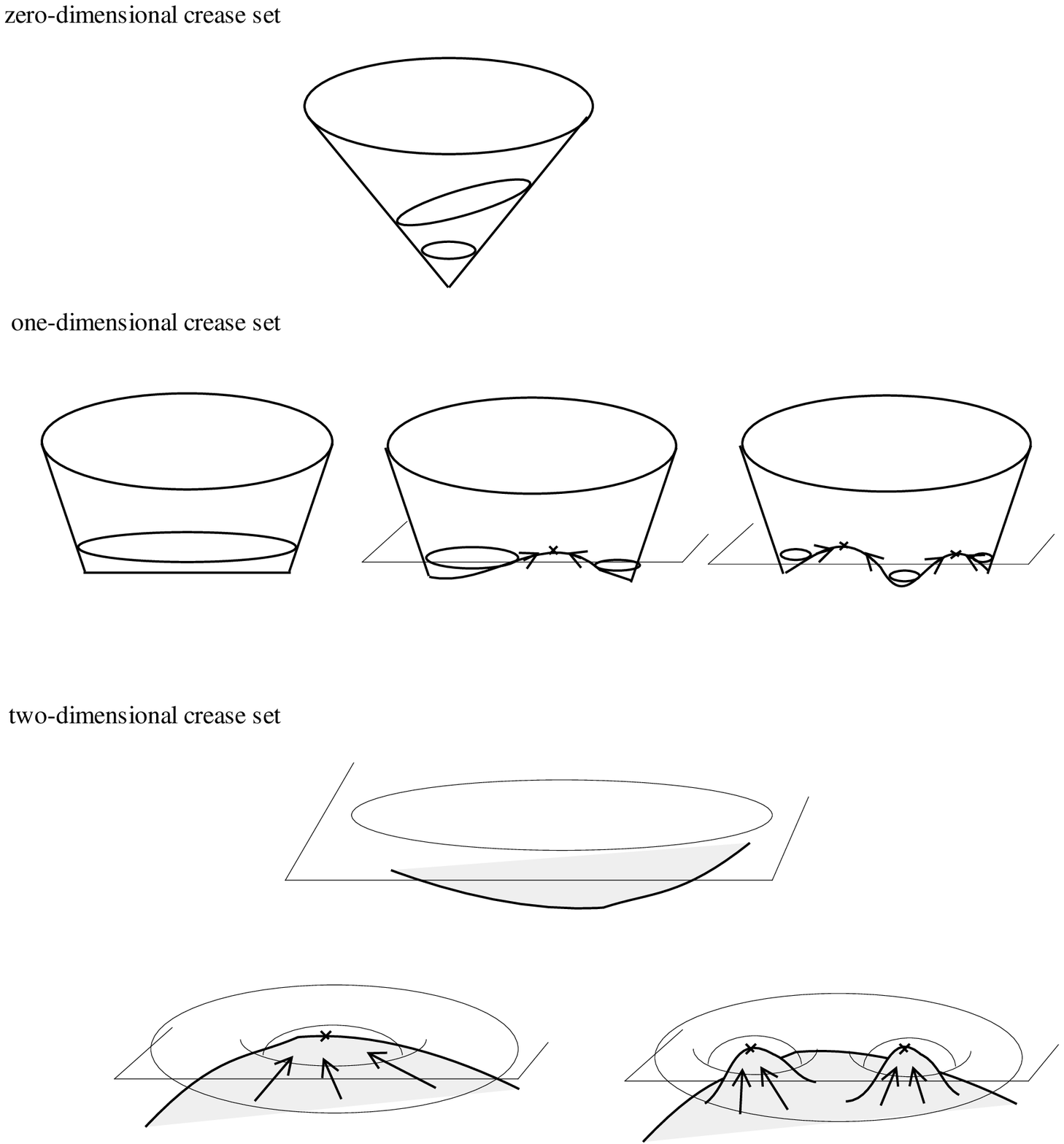}}
        \caption{EHs with the zero-, one- and two-dimensional crease sets are 
        shown. We see that the one-dimensional crease set becomes a coalescence 
        of an arbitrary number of spherical EHs. For the two-dimensional 
        crease set, only sections of the EH and the crease set are drawn. It can 
        become an EH with an arbitrary number of handles. It is also 
        possible to change the EH into a trivial creation of a spherical EH.}
        \protect\label{fig:ndend}
\end{figure}

A certain timeslicing gives the further changes of the Euler number. 
\begin{Corollary}\label{CL1}
An appropriate deformation of a timeslicing turns a process where the TOEH changes
from $n$ ($n=1,2,3,\cdots$) spheres to a sphere into a process where the TOEH 
changes from $m (m\neq n)$
spheres to a sphere. Also an appropriate deformation of a timeslicing turns 
a process where the TOEH changes
from a surface with genus = $n$ ($n=1,2,3,\cdots$) to a sphere into a process 
where the TOEH changes from a surface with genus = $m (m\neq n)$ to a sphere.
\end{Corollary}
\proof
From Theorem \ref{T2}, when the TOEH changes from $n \times S^2$ to a 
single $S^2$ in a timeslicing, there should be a one-dimensional crease set 
(in which there may be some branches).
 Since the crease set is an acausal set (Proposition \ref{P2}),
 there is another appropriate timeslicing hitting the crease set at $m$ 
 different points simultaneously (Fig.\ref{fig:ndend}(b)). On this timeslicing,
 the Euler number changes by $-2
   \times m$, and $m+1$ spheres coalesce. 
   Using the same logic, the EH of a surface with genus = $n$ can be regarded as
   the EH of a surface with genus = $m$
   by the appropriate change of its timeslicing (see Fig.~\ref{fig:ndend}(c)). \qed
   
As shown in Corollary \ref{CL1}, the TOEH depends strongly on the 
timeslicing. Nevertheless, Theorem \ref{T2} tells us that there is a 
difference
between the coalescence of $n$ spheres,  where the Euler number decreases by the
one-dimensional crease set, and the EH of a surface with genus = $n$, where the Euler number
increases by the two-dimensional crease set.

Finally we see that, in a sense, the TOEH is a transient term.
\begin{Corollary}
          All the changes of the TOEH are reduced to the trivial creation of 
          an EH which is topologically $S^2$.\footnote{As depicted in 
          Fig.\ref{fig:ndend}, this single black hole is expected to 
          be highly deformed.}
\end{Corollary}
\proof
We choose a point $p_c$ on the boundary of the crease set. Since the
collared crease set is topologically $D^3$ from Proposition \ref{P3}, by 
a distance function $l(p)=|p-p_c|$ along th crease set, we can slice the 
crease set by $l(p)=\rm const.$, and sections by this slicing do not
intersect each other. Moreover, because the crease set is an acausal
set, such a slicing of the crease set can be extended into the spacetime 
concerned as a timeslicing, so that $p_c$ becomes most in the past of
the crease set.
 In this timeslicing, since the crease set is sliced without degeneration, the zeros of $K$ appear only on the
 boundary of the crease set. Then, $\widetilde{H}$ has only one significant zero $p_c$ of
 $\overline{K}$ (type 1 in Fig.~\ref{fig:zeros}(b)), which corresponds to
 the point where the EH is formed, and meaningless zeros (with the index 
 0, for example, see Fig.~\ref{fig:ends}(c)) on the edge of the crease set. The index of $p_c$ is +1, and a 
 spherical EH is formed there. \qed

\begin{figure}
\centerline{\epsfxsize=14cm \epsfbox{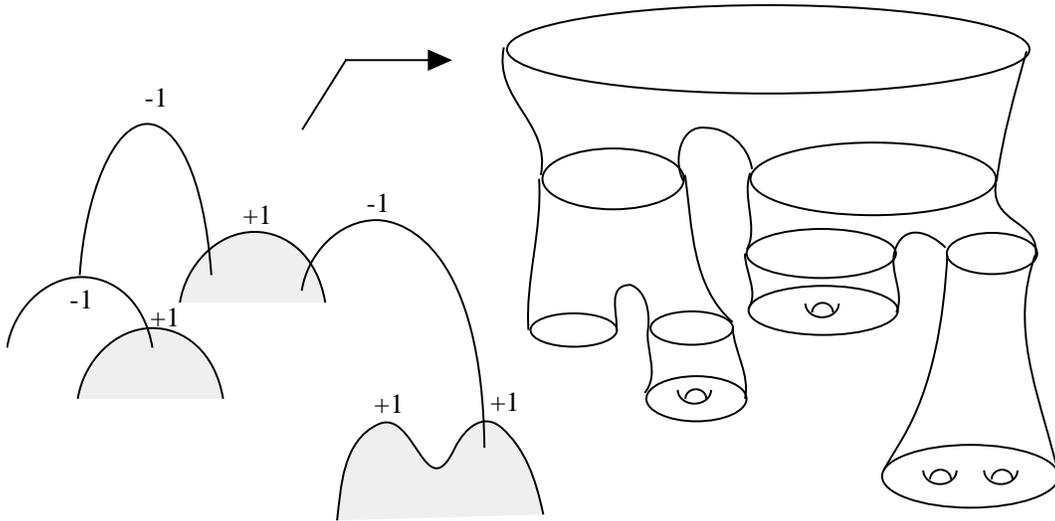}}
\caption{An example of the graph of the crease set is drawn. Determining the order of the 
vertices, we see the TOEH from the index of each zero.}
\label{fig:graph}
\end{figure}

Thus we see that the change of the TOEH is determined by the topology of 
the crease set and its timeslicing. For example, we can imagine the graph of the
crease set as Fig.~\ref{fig:graph}.
To determine the TOEH we must only give the order to each
 vertex of  the graph by a timeslicing. The graph in Fig.~\ref{fig:graph}
 may be rather complex. Nevertheless, considering a small scale 
  inhomogeneity, for example the scale of a single particle, the EH may admit
 such a complex crease set. It will be smoothed out in macroscopic physics.

\section{The linear perturbation of an event horizon 
with a spherical topology}
The purpose of this section is to investigate the stability of the TOEH 
which is always a sphere under linear perturbation. From 
our investigation in the previous section,\cite{MS1}
such an EH has only one zero-dimensional crease set (see Fig. \ref{fig:ndend}(a)). Then, we investigate whether this 
zero-dimensional crease set is stable under linear perturbation. Now, it
should be noted the `stability' does not imply that the perturbation does not blow
 up but that
the TOEH is not changed by the perturbation.

 As the
background spacetime, a spherically symmetric spacetime is appropriate. 
If the spherically symmetric spacetime has a non-eternal EH, it has only 
one endpoint at the origin, and the TOEH is always a sphere. In this 
case, it is possible to study the linear perturbation in an established.
framework\cite{SG}

Now we consider the Oppenheimer-Snyder spacetime as a familiar example 
of the EH with a spherical topology.
Its line element is given by

\begin{itemize}
        \item interior:
\begin{eqnarray}
        ds^2&=&a\left(\eta\right)^2\left(-d\eta^2+d\chi^2+\sin^2\chi 
        d\Omega^2\right),\ \ \ 0\le\chi\le\chi_0
        \label{eqn:int}\\
        a(\eta)&=&\frac12 a_m(1+\cos\eta),
\end{eqnarray} 
    \item       exterior:
\begin{eqnarray}
        ds^2 &=& -\left(1-2m/R\right)dt^2+{dR^2 \over 1-2m/R}+R^2 
        d\Omega^2,\ \ \ R_B(t)\le R ,\label{eqn:ext}\\
             &=& \left({32m^3\over R}\right)e^{-R/2m}\left(-dV^2+dU^2\right)
             +R^2 d\Omega^2, 
\end{eqnarray}
\end{itemize}
where $V$ and $U$ are Kruskal-Szekeres coordinates.
When these geometries are continued at $\chi=\chi_0$, the parameters of 
the exterior region are related to $a_m$ and $\chi_0$ as
\begin{eqnarray}
         m& = & \frac12 a_m\sin^3\chi_0,
        \label{} \\
         R_B& = & {a_m\sin\chi_0\over 2}(1+\cos\eta).
        \label{}
\end{eqnarray}

In the background spacetime the equations of null geodesics are easily 
solved and integrated. The background values of an outgoing null geodesic 
$\gamma$ in the direction $\theta_0, \phi_0$ and from the origin at $\eta=\eta_0$ are
\begin{eqnarray}
        l_0^a & = & \left({\partial \over \partial \eta}\right)^a+
        \left({\partial \over \partial \chi}\right)^a     \label{eqn:lin} \\
         & = & \left({\partial \over \partial V}\right)^a+
        \left({\partial \over \partial U}\right)^a       \label{eqn:lex} \\
        \gamma(\eta_0,\theta_0,\phi_0)&:&\ \ \ 
        \cases{\chi=\eta-\eta_0,\cr
        U-U_0(\chi=\chi_0,\eta=\chi_0+\eta_0)=
        V-V_0(\chi=\chi_0,\eta=\chi_0+\eta_0)\cr}\\
        \theta &=&\theta_0, \ \ \ \phi=\phi_0, \\
        \eta_{crit} & = & \pi-3\chi_0,
\end{eqnarray}
where $l_0$ is an outgoing null vector field and $\eta_{crit}$ is the 
supremum of the time $\eta$ when 
a light ray emitted from the origin can 
reach to the future null infinity $\scri^+$. The crease set of the EH is a point at 
the origin with $\eta=\eta_{crit}$. 

We treat the freedom of linear perturbation with a spherical harmonics 
$Y_{LM}$ expansion, and they are decomposed into odd 
parity $[(-1)^{L+1}]$ modes and even parity $[(-1)^{L}]$ modes.  Since they are decoupled in the spherically 
symmetric background, we discuss the stability of the TOEH under each mode 
of the perturbation with a 
parity, $L$, and $M$. First we develop the properties of null geodesics 
in a perturbed spacetime. The equations of null 
geodesics are given by
\begin{eqnarray}
        0 & = & g_{ab} l^a l^b 
        \label{} \\
          & = & \left(g_{0ab}+h_{ab}\right)\left(l_0^a+\delta 
          l^a\right)\left(l_0^b+\delta l^b\right)
        \label{} \\
          & = & h_{ab} l_0^al_0^b + 2 g_{0ab} \delta l^a l_0^b,
        \label{}
\end{eqnarray}
and
\begin{eqnarray}
        l^a\nabla_a l^b & = & \alpha l^b
        \label{} \\
        l^a\partial_a l^b +  \Gamma^b_{ac} l^al^c&=&\alpha l^b \rightarrow 
        \label{} \\
        l_0^a\partial_a \delta l^b +\delta l^a\partial_a l_0^b+2\Gamma^b_{0ac} l_0^a \delta l^c+\delta\Gamma^b_{ac}
         l_0^a l_0^c& =&\delta\alpha l_0^b+\alpha_0\delta l^b,  
        \label{}      
\end{eqnarray}
where $g_0, \Gamma_0$ is given by (\ref{eqn:int}), (\ref{eqn:ext}) and 
$l_0$ (\ref{eqn:lin}), (\ref{eqn:lex}).
the quantity $\delta\alpha$ corresponds to the parametrization of $l$, and it is set so 
that $\delta l^{\eta}+\delta l^{\chi}$ ($\delta l^V+\delta l^U$) vanishes. The deformation $\delta x^a$ of the light path $\gamma$ 
fixing its end on the same position of the future null infinity $\scri^+$
  is integrated backward along the 
background light path $\gamma$ from the future null infinity to a point 
in the interior region as
\begin{eqnarray}
     \delta\alpha&=&\delta\Gamma^{\eta(V)}_{ab}l_0^al_0^b+\delta\Gamma^{\chi(U)}_{ab}l_0^al_0^b\\
         \delta \eta&=&-\delta \chi=\int^{\chi}_{\chi=\chi_0
         }d\chi[\gamma]{h_{\eta\eta}+2h_{\eta\chi}+h_{\chi\chi}\over 4 
         a^2}+\delta\eta(\chi=\chi_0),  \label{eqn:dec}\\
         \delta\eta(\chi=\chi_0)&=&{\partial\eta\over\partial V}\delta 
         V_0+{\partial\eta\over\partial U}\delta U_0, \\ \delta V_0=-\delta U_0&=&
         \int^{U_0}_{U=\infty 
         }dU[\gamma]{(h_{VV}+2h_{VU}+h_{UU})Re^{R/2m}\over 128m^3},
        \label{}
\end{eqnarray}
since $\delta V_0=-\delta U_0$ implies $\delta\eta(\chi_0)=-\delta\chi(\chi_0)$.

\subsection{Even parity mode}
The metric perturbation of the even parity mode is given by
\begin{equation}
        h_{ab}=
        \bordermatrix{
        &
         \eta,V\  \chi,U
        &
                        \theta\ \ \    \phi 
                        \cr
        &\bar{h}_{AB}Y_{LM}  &  \bar{h}_AY_{LM,\alpha}\cr
        &       Sym    &  r^2(K\gamma_{\alpha\beta}Y_{LM}+GY_{LM;\alpha\beta})\cr
        },
        \label{eqn:eve}
\end{equation}
where $r$ is a circumference radius, and $\gamma_{\alpha\beta}$ is the 
metric of the 
unit sphere.\cite{SG}
For the even parity mode, the angular distribution of the $\delta\eta$ 
and  $\delta\chi$ (\ref{eqn:dec}) is 
given by the spherical harmonics $Y_{LM}$. So, it is helpful to discuss the symmetry 
of each $Y_{LM}$.

 Since $Y_{00}$ is a spherically symmetric 
function it causes no change of the crease set, unless
the perturbation is unstable and destroys the entire EH.
The even parity modes with $L=1$, $M =\pm 1$, can change into the mode of $Y_{10}$ 
through a certain rotation, and we only consider $M=0$ for the $L=1$ mode. By
a $Y_{10}$ perturbation, the wave front of light around 
the origin is shifted along the $z$-axis. Then, we only need to determine 
perturbed light paths starting from the origin and moving toward the north and the 
south. From Eq. (\ref{eqn:eve}), we see 
$\delta\chi(\gamma(\eta=\eta_0,\theta=0))=-\delta\eta(\gamma(\eta=\eta_0,\theta=0))=-\delta\chi(\gamma(\eta=\eta_0,\theta=\pi))=  
\delta\eta(\gamma(\eta=\eta_0,\theta=\pi))$.
Furthermore, $\delta\theta$ and $\delta\phi$ for these light paths vanish due to 
axial-symmetry. These imply the intersection of 
$\gamma(\eta=\eta_{crit},\theta=0)$ and 
$\gamma(\eta=\eta_{crit},\theta=\pi)$  does not undergo a change of its time $\eta$ 
but, rather, its position $\chi$ by $2\delta\chi$ along the $z$-axis. 
Since there is no peak of $Y_{10}$ between $\theta=0$ and $\theta=\pi$, all the 
other $\gamma(\eta=\eta_{crit},\theta,\phi)$ should also be shifted so as to pass 
the same position of $2\delta\chi$ on the $z$-axis at $\eta=\eta_{crit}$. Therefore the original 
zero-dimensional crease set is
only shifted in the $z$-direction by $2\delta\chi$. There is no change of 
the TOEH.

\begin{figure}
        \centerline{\epsfxsize=14cm \epsfbox{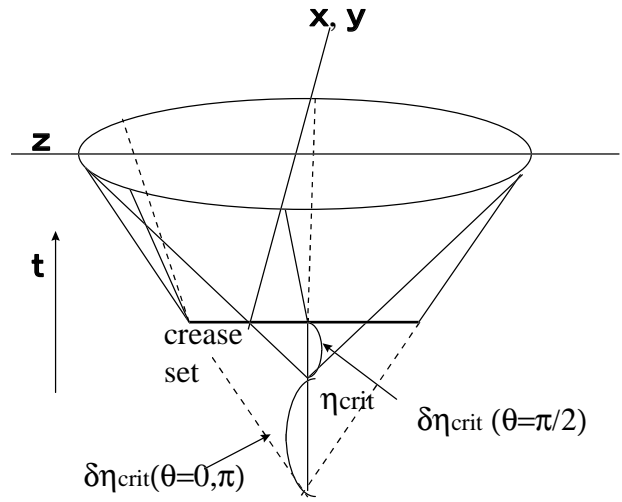}}
        \caption{The latest light paths with maximal $\delta\eta_{crit}$ ($x,y$
          direction in this figure) form an endpoint at the origin with
        $\eta=\eta_{crit}+\delta\eta_{crit}(\theta=\pi/2)$. On the other hand,
        a light path on the other axis ($z$-axis in this figure) crosses
        light paths from other directions and forms an endpoint
        there. Thus the crease set comes to possess a dimension in this ($z$-)direction.}
        \protect\label{fig:pend}
\end{figure}

The even parity mode with $L=2$ possesses reflection symmetries about three
orthogonal planes. For small perturbation, these modes change the 
wave front of light from spherical to ellipsoidal. By an appropriate rotation, the principal
axes of the ellipsoidal
wave front become the $x$-, $y$- and $z$-axis. Then, it is sufficient to determine
light paths along these axes. By the symmetry, $\delta\theta$ and
$\delta\phi$ vanish for these light paths. Since $\delta\eta=-\delta\chi$ 
implies that the change of $\eta_{crit}$ is given by 
$2\delta\eta(\gamma(\eta_{crit}))$, the $\delta\eta_{crit}$ of the light 
paths on the principal axes by each even 
parity $Y_{2M}$ mode are given by
\begin{eqnarray*}
  \delta\eta_{crit}(L=2,M=0)&=&\sqrt{\frac5\pi}H,
  -\frac12\sqrt{\frac5\pi}H,-\frac12\sqrt{\frac5\pi}H,\\
 \delta\eta_{crit}(L=2,M=1)&=&\frac32\sqrt{\frac5{6\pi}}H,
0, -\frac32\sqrt{\frac5{6\pi}}H,\\
 \delta\eta_{crit}(L=2,M=1)&=&\frac32\sqrt{\frac5{6\pi}}H,
0, -\frac32\sqrt{\frac5{6\pi}}H,
\end{eqnarray*}
where $H$ (the factor not depending on $Y_{LM}$) is given 
by Eq. (\ref{eqn:dec}).
From these results, we see the shape of the crease set around the origin. Light 
paths from the latest direction (maximal $\delta\eta_{crit}$) form an endpoint 
at the origin (for example, see Fig.~\ref{fig:pend}). On the other hand, light 
paths on the other axes 
cross a light path from another direction not passing through the origin, 
at a position different from the origin, so that their intersections provide the dimensions 
of the crease set to their directions.
Thus, the case of $L=2, M=\pm 1,2$ provides a two-dimensional crease set. 
Contrastingly the crease set with $L=2, M=0$ depends on the signature of $H$. If 
$H$ is negative (positive), the crease set is one (two)-dimensional (see Fig.~\ref{fig:pend}). Since $H$ 
is generally not equal to zero, the TOEH is not stable under the 
perturbation with $L=2$.

By the mode with
$L>2$, the wave front will experience more complicated deformation. By such a
deformation, the crease set will undergo branching and become highly 
complicated, as stated in the remark of Theorem\ref{T2}. For these 
modes,  $\delta\theta$ and $\delta\phi$ will not be
excluded from the discussion. A detailed investigation, however, would show the change of the
structure of the crease set occurs even with non-vanishing $\delta\theta$,
$\delta\phi$.

\subsection{Odd parity mode and higher order contribution}
The metric perturbation of the odd parity mode is given by
\begin{equation}
        h_{ab}=\bordermatrix{
                   &
                   \eta,V \  \chi,U 
                   &
                   \theta \ \ \ \phi 
                    \cr
                    &\begin{array}{cc}
                        0 & 0  \\
                        0 & 0
                    \end{array}  &  \bar{h}_AS_\alpha  \cr
                    & Sym         &  \bar{h}S_{(\alpha;\beta)}\cr
        },
        \label{eqn:odd}
\end{equation}
where the $S_\alpha$ are the transverse vector harmonics on the unit 
sphere.\cite{SG}
From (\ref{eqn:dec}) and (\ref{eqn:odd}), it is clear that the odd parity
mode does not affect $\delta\eta$, $\delta\chi$ at linear order. On
the other hand, though $\delta\theta$ and $\delta\phi$ exist, they do not
affect the structure of the crease set, because without $\delta\eta$ and
$\delta\chi$, all the perturbed outgoing light paths whose original past endpoint 
in background 
is the origin at $\eta=\eta_0$ start at the origin at the same time $\eta_0$. They
still have only one endpoint at the origin with $\eta=\eta_{crit}$. 

For the modes not changing the structure of the crease set, it would be 
necessary to investigate contributions from higher order evaluation. The higher order
contributions  are contained in the back-reaction of the changes of the
light path to the equation of null geodesics.
Nevertheless, it is also necessary to include a second order
metric-perturbation. It will cause difficulty in further
investigations.  At second order, there should be mode coupling between
different parities of  $L$ and $M$. This fact implies generally that the
structure of the crease set is unstable at higher order. Even if this is the case, however,
there are differences
in the sensitivity of the crease set among the modes. The TOEH is insensitive to
the odd parity mode and $L=1$ even parity mode.
\section{The structural stability of the topology of the event horizon}
In the previous section, it was shown that the spherical TOEH is unstable 
under linear perturbation. Since there is no appropriate example of
a spacetime, however, with non-spherical topology, similar analysis is
impossible for other TOEHs.
In this section we discuss the structural stability of the crease set of 
the EH in the context of catastrophe theory. As discussed in \S 2, this corresponds to the stability of the 
TOEH. First, we investigate this structural stability in a (2+1)-dimensional 
spacetime. 
\subsection{In (2+1)-dimensional spacetime}

The plan of analysis is following. First, we consider the appropriate wave
front of light in a flat spacetime. According to geometrical optics, 
the wave
front produces backward caustics and the endpoints of a null surface 
related to the wave front. In the context of catastrophe theory, Thom's
Theorem states that the structures of such caustics are 
classified\cite{PS} if they are 
structurally stable. Thus we analyze the structure of the caustics and judge 
whether the crease set is classified by Thom's Theorem. Here, we consider
the structural stability as corresponding to stability under the small change
of the shape of the wave front and the local geometry around the
endpoints. The shape of the wave front reflects the global structure of 
the spacetime between $\scri^+$ and the wave front. Furthermore, the stability is that of only the local structure 
of the caustics. Therefore, to discuss stability in a flat spacetime is valid as 
long as we deal with the structure of a small neighborhood.

\begin{figure}
        \centerline{\epsfxsize=16cm \epsfbox{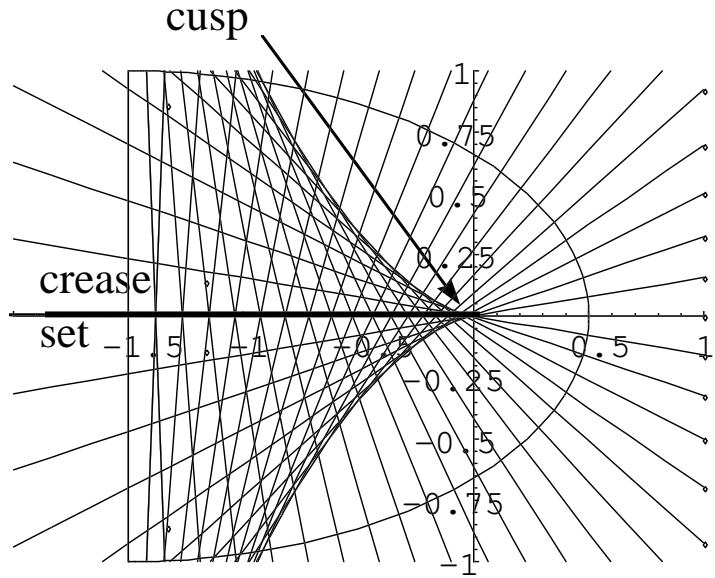}}
        \caption{The light paths for the elliptic wave front with $a=2,
          b=1$. 
        There are crossing points of the light paths which are the 
        endpoints of a null surface corresponding to the wave front, on 
        the $x$-axis, $[2x_0,0]$. A cusp is formed at the origin.}
        \protect\label{fig:cau2}
\end{figure}

For simplicity, we consider only an elliptical wave front, 
\begin{eqnarray}
               E_2:\  \left({x-x_0\over a}\right)^2+\left({y\over b}\right)^2=1\\
                x_0=-{a^2-b^2\over a},\ \ \  a\geq b\ .
        \label{}
\end{eqnarray}
Then, the square of the distance between ($x,y$) and ($X,Y$) is given by
\begin{eqnarray}
        f_{XY}\left({\bf x}\right)&=&\left(X-x\right)^2+\left(Y-y\right)^2\\
               &=&\left(X-\left(a \sqrt{1-\left(y/b\right)^2}+{b^2-a^2\over a}\right)\right)^2+\left(Y-y\right)^2,
        \label{}
\end{eqnarray}
where ($X,Y$) is an arbitrary point and $x-x_0$ is positive.
As known from geometrical optics, in a flat spacetime, a light path through
($X,Y$) is 
given by the stationary points of $f_{XY}\left(x,y\right)$:
\begin{eqnarray}
        {\partial f_{XY}\left({\bf x}\right)\over \partial y} & = & 0 \ \ \ \Rightarrow
        Y  =  A(y) X +B(y)
        \label{} \\
        A(y)&=&-\left({\partial x(y)\over \partial y}\right)_{E_2} , \ \ 
        B(y) = x\left({\partial x(y)\over \partial y}\right)_{E_2}+y,
        \label{}
\end{eqnarray}
where $(\partial/\partial)_{E_2}$ represents partial differentiation with the constraint $E_2$.
The light paths are drawn in Fig.~\ref{fig:cau2}. From this figure, we see 
that
they form a caustic at the origin, and the set of endpoints of the null surface 
concerning the wave front is a one-dimensional set, an interval on the 
$x$-axis  $[2x_0,0]$.

To see the structure of the caustic, we derive the Taylor 
series of $f_{XY}\left({\bf x}\right)$ around the origin,
\begin{eqnarray}
        \widetilde{f_{XY}}\left({\bf x}\right) & = & \left({b^4 \over a^2}-{2 b^2 X \over a}+X^2+Y^2\right)-
        2 Y y +{a X y^2 \over b^2} \\
         & + & \left({a^2 \over 4 b^4}-{1 \over 4 b^2}+{a X \over 4 
         b^4}\right)y^4+O\left(y^5\right).
        \label{}
\end{eqnarray}
Hence, the light paths form a cusp (a type $A_3$ catastrophe) at $(X,Y)=(0,0)$ 
because
$\widetilde{f}\sim y^4$. 
From Thom's Theorem, we know it is structurally stable except for at $a=b$. Of course, $a=b$ corresponds the
circular wave front and the zero-dimensional crease set.

\subsection{In (3+1)-dimensional spacetime}
\begin{figure}
        \centerline{\epsfxsize=16cm \epsfbox{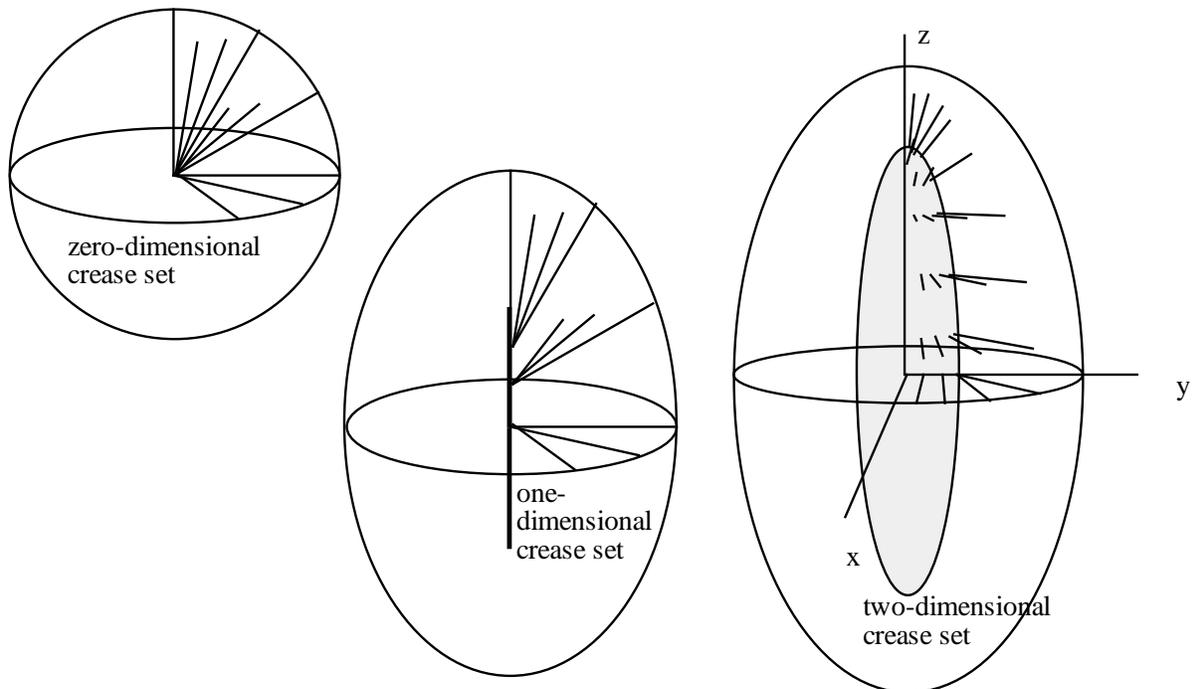}}
        \caption{The crease set becomes zero-dimensional for the spherical wave 
        front. In the prolate-spheroidal wave front, a one-dimensional 
        crease set appears. Otherwise, the ellipsoidal wave front produces a 
        two-dimensional crease set.}
        \protect\label{fig:eli}
\end{figure}

\begin{figure}
        \centerline{\epsfxsize=16cm \epsfbox{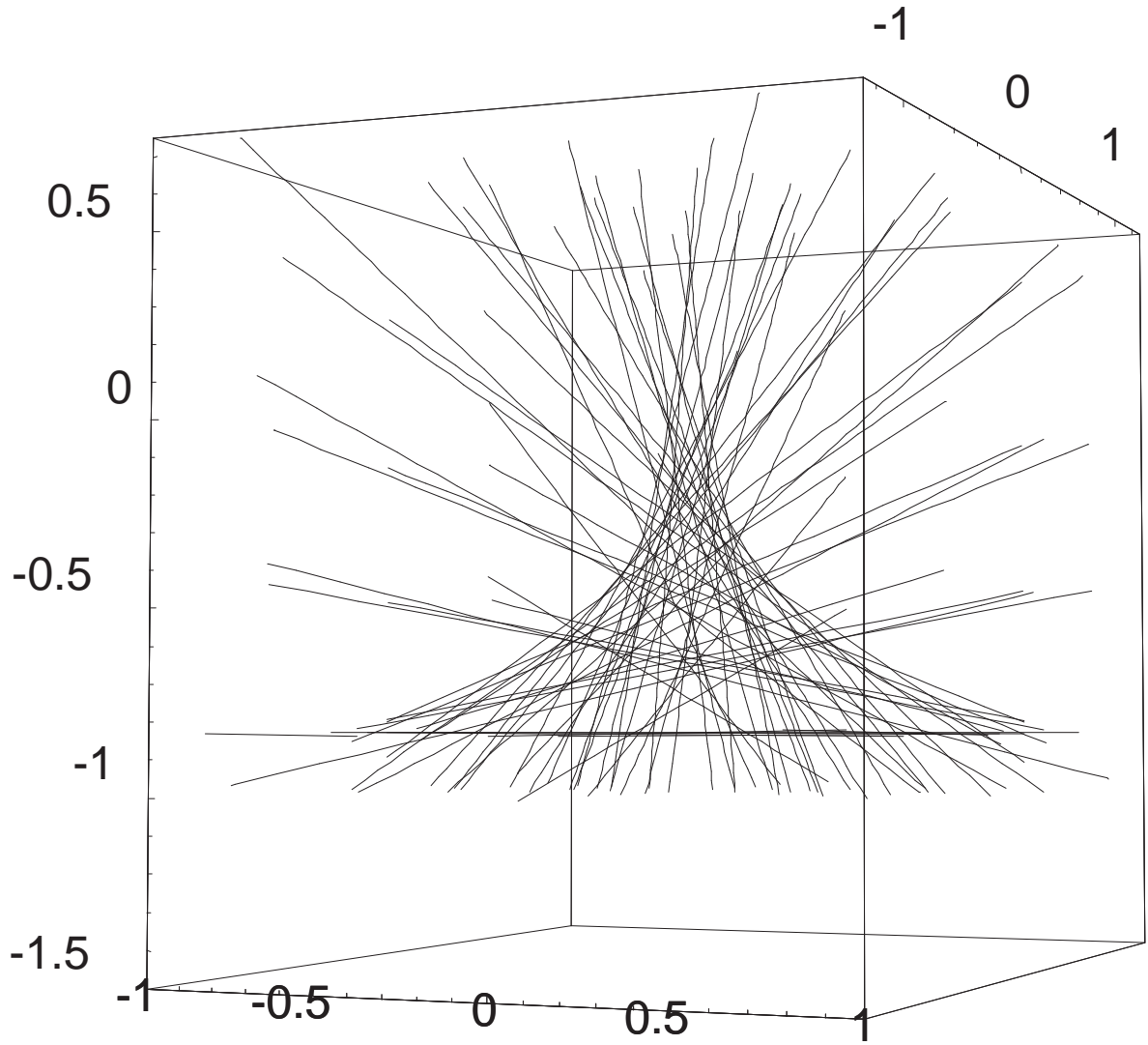}}
        \caption{The light paths for the ellipsoidal wave front with
          $a=1,\ b=1.3,\ c=1.5$.  A cusp is formed at the origin. Studying this figure 
        carefully, one sees that another cusp exists at $(0,0,(b^2-a^2)/c)$.}
        \protect\label{fig:cau3}
\end{figure}

For (3+1)-spacetime, the investigation above can
be carried out in a similar manner, though the situation becomes a little complex. In this case, 
there are 
three possibilities of the crease set of the EH, even after sufficient 
simplification. As shown in Fig.~\ref{fig:eli}, the endpoint forms a 
point, line, or surface. As in the previous subsection, we consider the ellipsoidal
wave front,
\begin{eqnarray}
              E_3:  \left({x\over a}\right)^2+\left({y\over b}\right)^2+\left({z-z_0\over c}\right)^2=1\\
                z_0=-{c^2-a^2\over c},\ \ \  0< a\leq b\leq c.
        \label{}
\end{eqnarray}
For a $z-z_0>0$ branch, the square of the distance between $(x,y,z)$ and 
an arbitrary point $(X,Y,Z)$ is given by
\begin{eqnarray}
        f_{XYZ}\left({\bf x}\right)&=&\left(X-x\right)^2+
        \left(Y-y\right)^2+\left(Z-z\right)^2\\
               &=&\left(X-x\right)^2+\left(Y-y\right)^2+
               \left(Z-(c\sqrt{1-\left(x/a\right)^2-\left(y/b\right)^2}+
               z_0)\right)^2.
        \label{}
\end{eqnarray}
The light path through $(X,Y,Z)$ is given by
\begin{eqnarray}
        {\partial f_{XYZ}\left({\bf x}\right)\over \partial x} & = & 0,\ \ 
        {\partial f_{XYZ}\left({\bf x}\right)\over \partial y} = 0 
        \label{} \\
        \Rightarrow X & = & A(x,y) Z +B(x,y),\ \ \  and\ \   Y=C(x,y) 
        Z+D(x,y),
        \label{} \\
        A&=&-\left({\partial z(x,y)\over\partial x}\right)_{E_3} , \ \ 
        B =z\left({\partial z(x,y)\over\partial x}\right)_{E_3}+x,\ \  \\
        C&=&-\left({\partial z(x,y)\over\partial y}\right)_{E_3},\ \  
        D=z\left({\partial z(x,y)\over\partial y}\right)_{E_3}+y.
        \label{}
\end{eqnarray}
From Fig.~\ref{fig:cau3}, exhibiting the light paths, it is seen that a caustic  
is formed around
 the origin. Only when $a$, $b$ and $c$ are equal the crease set
 becomes
zero-dimensional (at the origin). The relations $a=b\neq c$ imply that endpoints form
a one-dimensional set which is an interval on the $z$-axis,
 $[2z_0,0]$. Otherwise, the crease set is two-dimensional (Fig.~\ref{fig:eli}).

The Taylor series of the potential $f$ at the origin is given by
\begin{eqnarray}
        \widetilde{f_{\bf X=0}}\left({\bf x}\right) &=& {a^4\over c^2}+
        {-a^2+c^2\over 4 a^4}x^4+
{-a^2+c^2\over 8 a^6}x^6
          +  {b^2-a^2\over b^2}y^2+{-a^2+c^2\over 2 a^2 b^2}x^2 y^2\\ &+&
         {3\left(-a^2+c^2\right)\over 8a^4 b^2}x^4y^2 + {-a^2+c^2\over 
         4 b^4}y^4+{3\left(-a^2+c^2\right)\over 8a^2b^4}x^2y^4+{-a^2+c^2\over 
         8b^6}y^6\\ &+& O\left({\bf x}^7\right).
        \label{}
\end{eqnarray}
The structure of the caustic is controlled by the leading term of
$\widetilde{f}$ about $x,y$. 
When $a<b\le c$, $\widetilde{f}\sim\alpha x^4 + \beta y^2+\gamma$ produces a cusp (type $A_3$).
Then, the two-dimensional crease set is structurally stable. On the other hand,
 $\widetilde{f}$ becomes 
$\alpha (y^4+2x^2y^2+x^4)+\gamma$ with $a=b\neq c$. This case corresponds to 
the line crease set and it is 
not structurally stable. Incidentally, if $a<b<c$, there is also another 
cusp at $(0,0,-(b^2-a^2)/c)$ (as seen by carefully studying Fig.~\ref{fig:cau3}). The 
Taylor expansion of $f$ around this cusp is given by
\begin{eqnarray}
        \widetilde{f_{\bf X=X'}}\left({\bf x}\right) &=& {b^4\over c^2}+
        {-b^2+c^2\over 4 b^4}y^4+
{-b^2+c^2\over 8 b^6}y^6
          +  {a^2-b^2\over a^2}x^2+{-b^2+c^2\over 2 a^2 b^2}x^2 y^2\\ &+&
         {3\left(-b^2+c^2\right)\over 8a^2 b^4}x^2y^4 + {-b^2+c^2\over 
         4 a^4}x^4+{3\left(-b^2+c^2\right)\over 8a^4b^2}x^4y^2+{-b^2+c^2\over 
         8a^6}x^6\\ &+& O\left({\bf x}^7\right).
        \label{}
\end{eqnarray}

This implies that the cusp is also stable as 
long as $a\neq b$ and $b\neq c$. With $b\rightarrow a$, this cusp 
approaches the cusp at the origin, and they degenerate into an unstable 
structure. Contrastingly, when $b$ is equal to $c$, this cusp 
disappears at the center of the ellipsoid. The example given in Ref.\cite{ST} 
corresponds to this case. Of course, the zero-dimensional crease set $(a=b=c)$  
is not structurally stable.

\section{An example and conjecture}
Figure \ref{fig:ndend}, and the discussion in the section 4 and 5 suggest that the TOEH is 
affected by the distortion of the black hole. Then, qualitative 
discussion is possible in regard to the relation between the TOEH and 
collapsing matter.
\subsection{The hoop conjecture and KT-solution}
First, we discuss the hoop conjecture in the sense of the TOEH, analyzing 
the EH of the Kastor-Traschen (KT) solution.\cite{ID}
The KT solution is the only known exact solution 
describing the coalescence of black holes. Since its EH is an
eternal one, it  
does not satisfy the assumption of \S 4. Nevertheless, it can 
be stated that the crease set is acausal and connected (not compact) also 
in the case considered here (which is the coalescence of two black 
holes with identical mass parameters). This fact leads us to relate the
KT  solution to the hoop 
conjecture.

On the grounds of physical considerations, one expects the formation 
of a black hole to occur when the matter or inhomogeneity of  
spacetime itself (non-linear gravitational waves) 
is concentrated into a sufficiently ``small" region. 
This might be the basis of the so-called 
{\it hoop conjecture} (HC) proposed by Thorne\cite{TH} 
which states that {\it black holes with 
horizons form when and only when a mass $M$ gets compacted 
into a region whose circumference in every direction is 
$C\leq 4\pi M$}. We point out here, as mentioned in \S 4 that the
coalescence of a black hole can be regarded as the formation of a
black hole which is expected to be highly distorted.
Hence, if we adopt a timeslicing in which the 
two stars {\it instantaneously} form a single, highly distorted black hole, 
the length scale of the black hole on this hypersurface 
is expected to be of the 
order of the Schwarzschild radius determined by its gravitational 
mass. Thus, the discussion in this subsection shows that the KT spacetime provides 
a good example, at the intuitive level, for this conjecture, 
at least as formulated for event horizons.

Since there is a positive cosmological constant, $\Lambda$,  
in the KT spacetime, this spacetime is not asymptotically 
flat but, rather, asymptotically de Sitter. 
Although the hoop conjecture has been proposed for an 
asymptotically flat spacetime, if the mass scale of black holes 
is much smaller than the cosmological horizon scale, 
$\sqrt{3/\Lambda}$, this conjecture should hold also in 
an asymptotically de Sitter spacetime. 
Furthermore it has been suggested that the hoop conjecture holds 
even for the case of gravitational collapse with a large 
mass scale in such an asymptotically de Sitter 
spacetime.\cite{Ref:CM}

Now, we focus on the KT spacetime of 
two black holes with identical mass parameters. 
The line element and the electromagnetic potential one-form 
in the contracting cosmological chart are given in the form
\begin{eqnarray}
ds^2&=&-U^{-2}d\tau ^2+U^2(dx^2+dy^2+dz^2),\\
A&=&-U^{-1}d\tau,
\label{eqn:cha}
\end{eqnarray}
where
\begin{eqnarray}
U&=&-H\tau+\frac{m}{r_+}+\frac{m}{r_-},\\
r_{\pm}&=&\sqrt{x^2+y^2+(z\mp  a)^2},
\end{eqnarray}
and $H$, $m$ and $a$ are constant. 
The constant $H$ is related to the cosmological constant via
$H=\sqrt{\Lambda /3}$. The above line element 
represents two point sources, each with mass 
parameter $m$ and charge $|e|=m$,
located at ${\bf x}=(0,0,\pm a)$
in the background Euclidean 3-space. 
The AD mass, $M_{tot}$, of this system is given by $2m$, which 
must be less than the critical value
$M_{crit}=1/4H$, so that the spacetime describes 
the merging of two black holes. Then its EH 
is located in the range $\tau <0$. With regard to the region of the spacetime 
which is covered by the above cosmological chart, see Refs.\cite{BHKT} 
and \cite{NSH}.

We numerically integrated equations 
for the null geodesics backward in time 
from a sufficiently late time $\tau =-\epsilon$, where $\epsilon$ is a 
sufficiently small positive parameter. 
In the contracting cosmological chart (\ref{eqn:cha}), 
the marginally trapped surface enclosing both black holes 
goes to infinity, $r\equiv\sqrt{x^{2}+y^{2}+z^{2}}\rightarrow 
+\infty$, in the limit $\epsilon \rightarrow 0$.\cite{NSH}
Hence, at $\tau=-\epsilon$, the marginally trapped surface 
enclosing both holes is sufficiently spherical, since 
the higher multipole moments are much smaller than the monopole 
term in the line element (\ref{eqn:cha}) in the limit $r\rightarrow+\infty$. 
The vicinity of the marginally 
trapped surface is well approximated 
by the one-black-hole solution, i.e., the Reisner-Nordstr\"om-de 
Sitter (RNdS) spacetime. 
Moreover, for this reason, the EH is very close 
to the marginally trapped surface enclosing both holes 
at $\tau=-\epsilon$. 
Therefore we integrate the equations for the past directed ingoing 
null geodesics normal to the marginally trapped sphere at
$\tau=-\epsilon$,
 which is easily
derived as
\begin{equation}
r=\frac{(1-\sqrt{1-8Hm})/2H-2m}{H\epsilon}.
\label{eqn:sph}
\end{equation}
Here the {\it ingoing} direction is determined in the usual sense 
with respect to the radial coordinate $r$. 
The above expression agrees with the location of the 
event horizon in the RNdS solution with mass parameter 
$2m$.\cite{BH} 
If we choose sufficiently small $\epsilon$, the deviation 
of the spacelike two-sphere defined by Eq.(\ref{eqn:sph}) from 
the EH should be smaller than the error due to the numerical 
integration. 

\begin{figure}
        \centerline{\epsfxsize=14cm \epsfbox{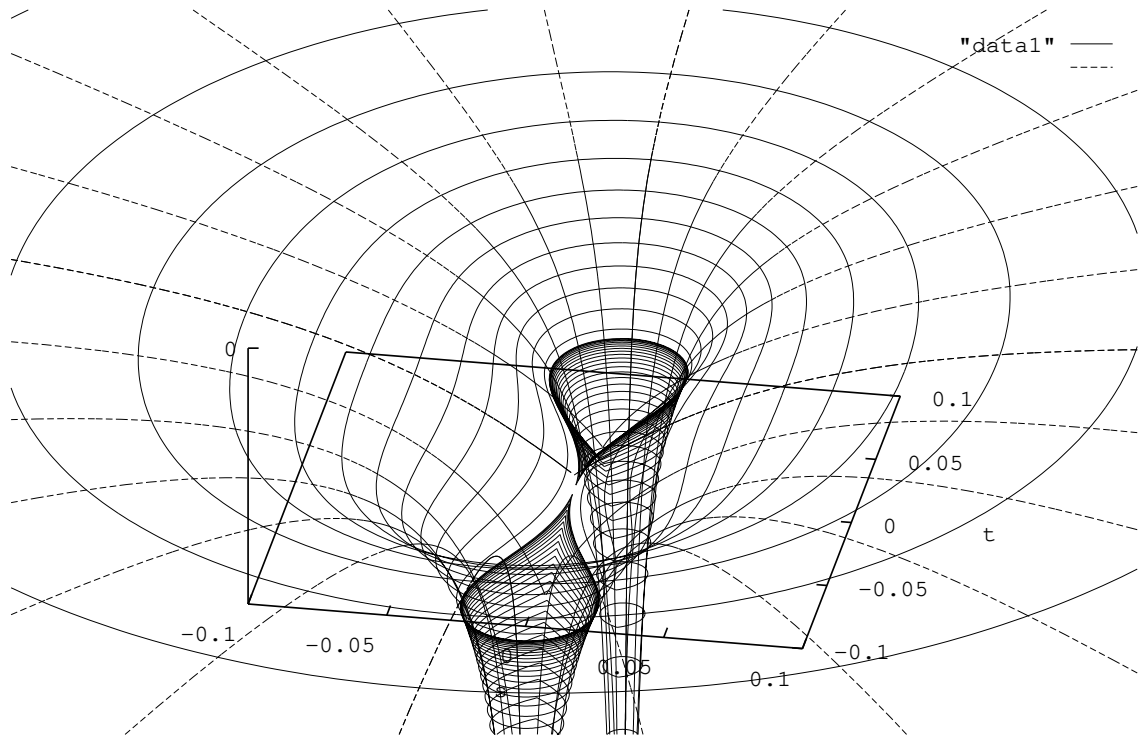}}
        \caption{The EH of the KT solution. It can be seen that 
          the two EHs prior to their coalescence are not smooth at the crease set 
        of the EH.}
        \protect\label{fig:ida1}
\end{figure}

Figure \ref{fig:ida1} depicts the time evolution of the EH 
in the contracting cosmological chart (\ref{eqn:cha}), 
which shows the usual picture of a two-black-hole collision. 
Adopting the normalization $H=1$,  
Fig.~\ref{fig:ida1} corresponds to the case with parameters 
$m=0.1$ and $a=0.1$. Then the AD mass of this system 
is $M_{tot}=0.8M_{crit}$. 
As can be seen from this figure, the crease set 
of the EH is a curve. 
In order to investigate whether or not the crease set is acausal, 
we have calculated the quantity
\begin{equation}
N\equiv g_{\mu\nu}m^{\mu}m^{\nu},
\end{equation}
where $g_{\mu\nu}$ is the metric tensor which gives the line element 
(\ref{eqn:cha}) and $m^{\mu}$ is the tangent vector of the crease set. 
With this analysis it is confirmed that the crease set is a spacelike
one-dimensional curve.  

\begin{figure}
        \centerline{\epsfxsize=14cm \epsfbox{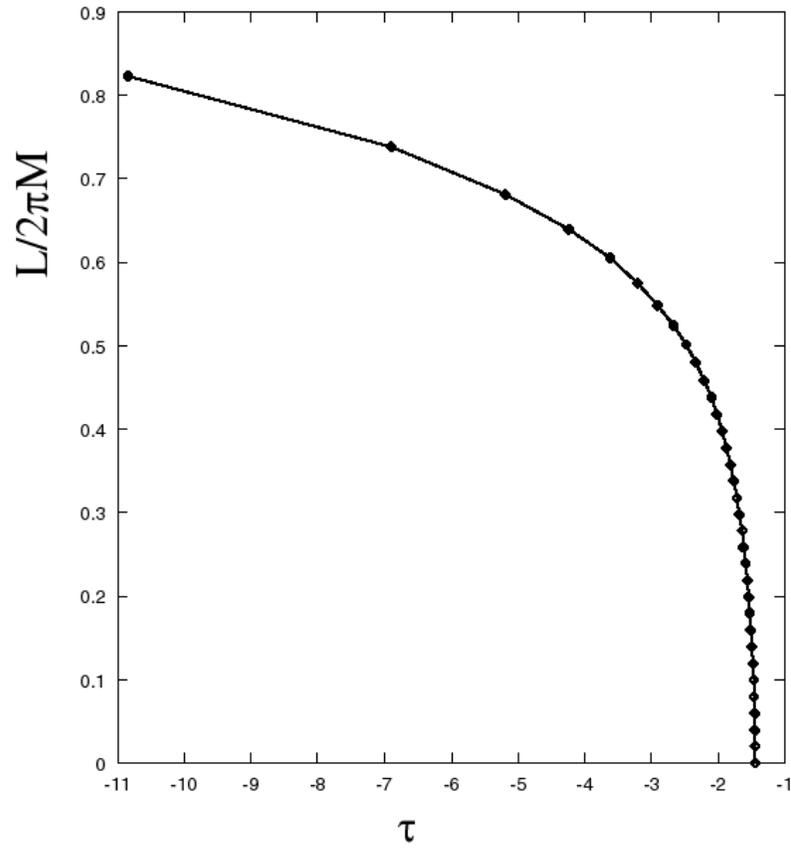}}
        \caption{The time $\tau$ dependence of the length of the crease set
          curve later than the time $\tau$. It seems that the length
        converges to $2 \pi M$ as $\tau$ goes to $-\infty$.}
        \protect\label{fig:ida3}
\end{figure}

In Fig.~\ref{fig:ida3} we give the numerical result for the proper length of the 
crease set, which is related to a spindle EH,
as a function of $\tau$. 
This figure shows that 
the length of the crease set converges on $\pi M_{tot}$ from below, while
the crease set is non-compact. This implies that there exists a distorted hoop
surrounding the crease set whose circumference never exceeds $4\pi M_{tot}$.

\begin{figure}
        \centerline{\epsfxsize=16cm \epsfbox{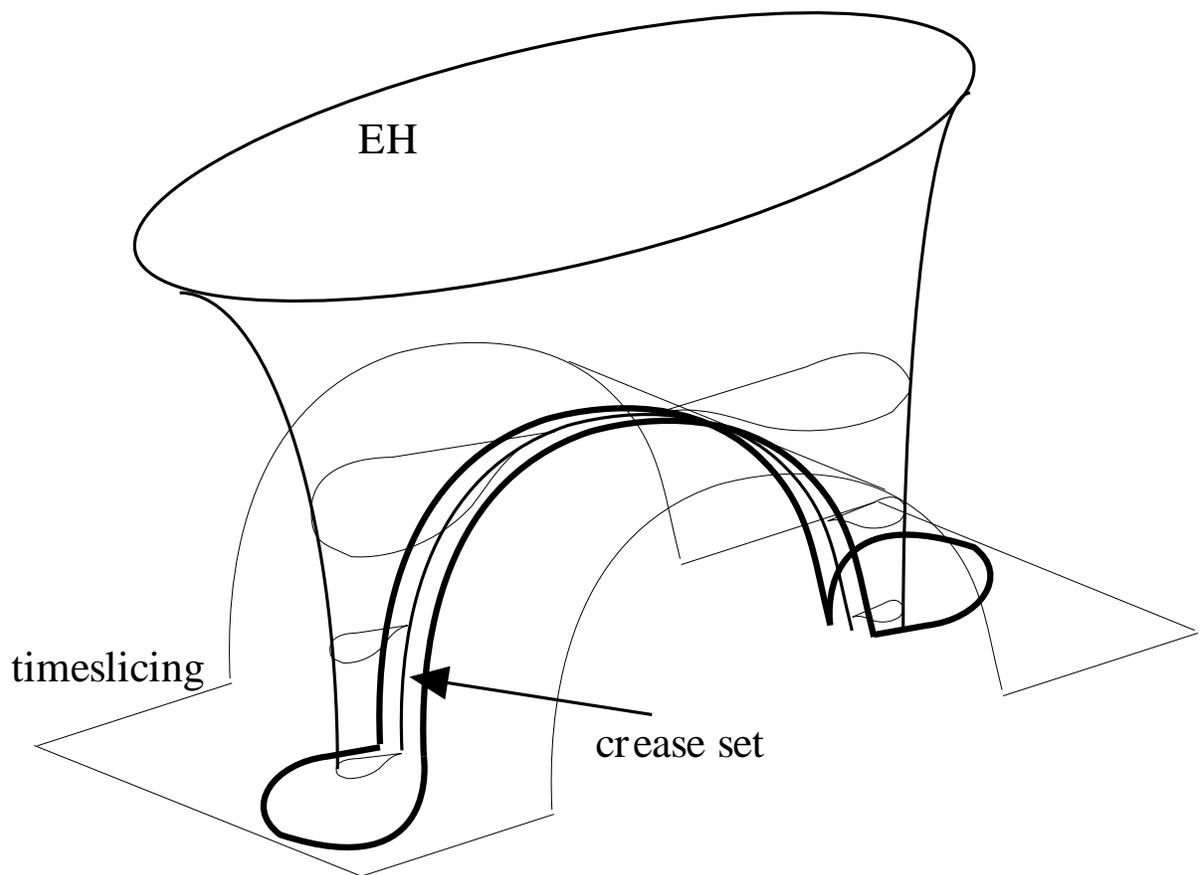}}
        \caption{The EH of the KT solution and a timeslicing. With this timeslicing the EH is regarded as one highly
        distorted black hole. It is surrounded by a hoop (drawn by a
        bold curve) whose circumference is $4\pi M$.}
        \protect\label{fig:ida4}
\end{figure}

Although the KT solution considered here describes an 
eternal black hole spacetime, we regard this 
spacetime as the far future of two stars composed of charged 
fluid.  
In the ``past" of this picture, the physical distance on the cosmological
chart (\ref{eqn:cha}) between each star is sufficiently large. Thus these
stars  should collapse
and be enclosed
by their marginally trapped surfaces, which do not enclose both stars but, rather, 
only one of the two. 
Let us consider such a case. Since the crease set is a spacelike curve, 
we can always adopt a timeslicing which includes a spacelike 
hypersurface, as shown in Fig.~\ref{fig:ida4}; the shape of a section 
of the event horizon for this spacelike hypersurface 
is dumbbell-like. 
Our result that the length of the crease set of the KT solution converges 
on
$2 \pi M_{tot}$
suggests that there should not be the
case in which the 
circumference of the EH produced from two massive charged
 stars is longer 
than order $4 \pi M_{tot}$ by such an appropriate choice of the 
timeslicing, even if the initial separation 
in the usual picture is much larger than order $M_{tot}$. 
We should note that 
there exist marginally trapped surfaces which enclose each 
hole (star) and will be apparent horizons, inside that EH.\cite{NSH}
This implies that the circumference of a black hole with (apparent) 
horizons is just bound by $4\pi M_{tot}$ in general.  
In this sense, the hoop conjecture seems to hold even in such a
spacetime including a cosmological constant. Here it should be noted
that the result presented here is only a tentative one. More detailed
and confirmed results will appear in our forthcoming paper.\cite{ID}

\subsection{Speculation: Scale conjecture of the TOEH}

One may expect to give some restriction on the TOEH 
introducing certain conditions on the matter field.
The present results in the section 4,5,6 imply, however, that this is probably hopeless. It seems that the symmetry of the spacetime affects
the structure of the crease set of the EH and the TOEH, and it is
easily disturbed by perturbation.
If one is not concerned with the scale of the topological structure of the EH, the TOEH 
can generally become complicated.
On the other hand, the discussion is the previous subsection is based on
the expected relation between the scale of the TOEH and the deformation 
of the black hole, where the scale dependent structure of the topology would 
be defined in a certain manner, for example Ref. \cite{SE}.
Here, taking the standpoint that we believe the existence of 
something like a ``hoop theorem'' in a wide class of gravitational
collapses, we also conjecture that the TOEH cannot be non-spherical 
in the sense of a smoothed topology (if it can be well defined) in its
mass scale without a naked singularity because, if we observe 
the non-spherical TOEH with a determination in its mass scale, 
the size of the crease set should be larger than the mass scale. The hoop 
theorem, however, will not admit such a deformed collapse that the crease set has 
the length of the mass scale, without the naked singularity.
\begin{Conjecture}({\bf The hoop theorem of the TOEH})
Smoothing the topological structure of an EH by its mass scale, the EH 
whose (spatial) topology is not a single sphere cannot be formed without 
a naked singularity.
\end{Conjecture}

Here, we must note that there is the possibility of a wide crease set 
without the corresponding configuration of matter, since the structure of 
the EH can also be determined by the asymptotic structure of a spacetime. Such a possibility 
would have to be eliminated by a certain asymptotic condition.
Assuming the hoop theorem,\cite{SY}\cite{FL} it might not be so difficult 
to construct the theorem conjectured here.

\section{Summary and discussion}
We have studied the topology of the EH (TOEH), partially considering the 
indifferentiability of
the EH. We have found that the coalescence of EHs is related to the one-dimensional
 crease set and a torus EH is related to the two-dimensional crease set. In 
 a sense, this is a 
 generalization of the result of Shapiro, Teukolsky, and 
 Winicour.\cite{ST}  Furthermore these
changes of the TOEH can be removed by an appropriate timeslicing,
since the crease set of an EH is a connected acausal set. We see that
 the TOEH 
depends strongly on the timeslicing. The dimension of the crease set,
however, plays an important role for the TOEH and, of course, is invariant under the change of the 
timeslicing. Then, a question arises, which dimension is 
generally possible for the crease set. 

Following these results, we have investigated the stability of the topology of the EH (TOEH).
First the stability of a spherical topology was
investigated under linear perturbation in a spherically symmetric
background. To linear order, the $L=2$ even parity mode changes the
structure of the crease set and the TOEH, and the odd parity mode and $L=1$ mode
do not. At higher order, however, mode coupling between the  modes with 
different $L$ or parity  cause the
instability of the TOEH even for these modes not changing the TOEH at the 
linear order. For $L> 2$ even parity modes, more detailed
investigation is required. In any case, we have seen that the trivial TOEH
is generally unstable under linear perturbation. 
 
In this discussion of the linear perturbation, we consider the
Oppenheimer-Snyder spacetime as an example of an always spherical
EH. Nevertheless, the result would be same for other non-eternal EHs with
spherical symmetry, since we have never used a concrete geometry of
the spacetime, other than the spherical symmetry.

How can we interpret the fact that the TOEH is insensitive to some
modes of the perturbation? In a sense, when we give an odd parity or $L=1$
 perturbation to the spherical EH, the
change of the TOEH at higher order would not be detectable
(though it is not trivial how one can observe it). This follows from the face that, while a local geometry around an observer is perturbed with
the same strength as the given perturbation,
the TOEH is not so perturbed. The change of the TOEH at higher order would be
prevented from being observed.

Next, using
 simple analysis from catastrophe theory, the structural stability of 
the crease set was 
studied in a more general situation. Assuming an ellipsoidal wave front, 
the stability of zero-, one-, and two-dimensional crease sets were
investigated. We see that the two-dimensional crease set is stable, and the
one- and zero-dimensional crease sets are not. Therefore, TOEHs with handles
(torus, double torus, ...) are generic.

Though in the present article we find only the crease set with a cusp 
catastrophe, as discussed
 in Ref. \cite{AR}, there 
is the possibility of additional types, the `swallowtail', the
`pyramid', and so on. They should form other structurally stable
crease sets. The TOEH in these cases them will be studied in a forthcoming work.

As an example of such a change of the TOEH, the KT solution is examined in 
its causal structure. Determining the EH of the KT solution describing 
the collision of two black holes with identical mass numerically, we 
see that the length of its crease set converges on a finite value $\pi
M_{tot}$ from below, while the crease set is non-compact. As mentioned in the
\S 4,  
 the crease set is a connected acausal set, and we can relate it to the hoop 
conjecture, because, since such an EH can be regarded as the formation of a
distorted black hole in a certain timeslicing. Then the crease set with its
length being $2 \pi M_{tot}$ implies the existence of a hoop whose
circumference is $4\pi M_{tot}$. In this sense, 
the bounds for the length of the crease set in the KT solution support the
hoop conjecture. 

Finally a conjecture about the scale of the TOEH was given. Assuming the 
``hoop theorem" we expect that only topological structure (e.g. 
handles) smaller than its mass scale is admitted without naked
singularity. After the definition of a scale-dependent topological
structure, this conjecture could be proved in the situation where a
restricted hoop theorem holds.

Incidentally, some of the statements in this article may be equivalent to  results 
of previous works.\cite{HA}$\sim$\cite{BG} Nevertheless the condition required here is 
quite different from that appearing in their works (for example,  energy conditions 
have never been assumed here). The present results may be considered as the
extension of those in the previous works.

Finally we are reminded of an essential question. How can we see the topology of the EH?
Some of the previous works, for example ``topological censorship'',\cite{FS} stresses 
that it is impossible.
On the contrary, we expect phenomena depending strongly on the existence of the EH as 
the boundary condition of fields, for 
instance the quasi-normal mode of gravitational waves\cite{CH} or Hawking 
radiation,\cite{HR} reflects the TOEH. For example, with regard to Hawking 
radiation, we would like to construct a toy model for the change of the TOEH, 
something like the Rindler spacetime for the Schwarzschild spacetime. 
This is our future problem.

\centerline{\bf Acknowledgements}

I would like to thank Professor H. Kodama, Professor J. Yokoyama, Dr. K. Nakao, Dr. S. Hayward, Dr. T. Chiba, Dr. 
A. Ishibashi and Dr. D. Ida
for helpful discussions. I am grateful to Professor H. Sato and 
Professor N. Sugiyama for their continuous encouragement.
I thank the Japan Society for the 
Promotion of Science for financial support. This work was supported in 
part 
by the Japanese Grant-in-Aid for Scientific Research Fund from the Ministry 
of 
Education, Science, Culture and Sports.


\end{document}